\documentclass[pra, preprint, superscriptaddress, showpacs]{revtex4}
\usepackage[nohyperlinks,smaller,printonlyused]{acronym}
\usepackage[english]{babel}
\usepackage{multirow}
\usepackage{booktabs} 	
\usepackage{bbold}
\usepackage{graphicx}
\usepackage{mathrsfs} 			
\usepackage{amsmath,amssymb,amsfonts,tikz}  
\usetikzlibrary{matrix}
\usepackage{xcolor}   		
\usepackage{epsfig}
\usepackage[pdftex]{hyperref}
\definecolor{MyDarkGreen}{rgb}{0,0.6,0}
\definecolor{MyDarkBlue}{rgb}{0.1,0.3,0.65}
\definecolor{MyDarkRed}{rgb}{0.7,0,0.15}
\definecolor{halfgrey}{rgb}{.8,.8,.8}

\DeclareMathOperator{\eul}{\mathrm{e}}

\newcommand{\dd}{\mathrm{d}}			\renewcommand{\phi}{\varphi}

\def\comment#1{}

					\def\pa{\partial}

							\def\s{\sigma}
						\def\t{\tau}						
												\def\p{\pi}
\def\a{\alpha}											\def\b{\beta}
																	
\def\g{\gamma}						\def\G{\Gamma}					\def\D{\Delta}
\def\w{\omega}						\def\W{\Omega}					\def\Y{\Psi}
											
\def\nn{\nonumber}										
											
											\def\x5{x^{5}}

											\def\fr{\frac}
\begin{document}

\title{Nonlinear effects in photoionization over a broad photon-energy range within the \acs{TDCIS} scheme}
\author{Antonia Karamatskou}
\email{antonia.karamatskou@cfel.de}
\affiliation{The Hamburg Centre for Ultrafast Imaging, Universit\"at Hamburg, Luruper Chaussee 149, D-22761 Hamburg, Germany}
\affiliation{Center for Free-Electron Laser Science, DESY, Notkestrasse 85, D-22607 Hamburg, Germany}

\begin{abstract}
The present tutorial provides an overview of the time-dependent configuration interaction singles (\acs{TDCIS}) scheme applied to nonlinear ionization over a broad photon-energy range. The efficient propagation of the wave function and the calculation of photoelectron spectra within this approach are described and demonstrated in various applications. Above-threshold ionization of argon and xenon in the extreme ultraviolet energy range is investigated as an example. A particular focus is put on the xenon $4d$ giant dipole resonance and the information that nonlinear ionization can provide about resonance substructure. 
Furthermore, above-threshold ionization is studied in the x-ray regime and the intensity 
regime, at which multiphoton ionization starts to play a role at hard x-ray photon energies, is identified. 
\end{abstract}

\pacs{Photoionization, multiphoton ionization, photoelectron spectra, strong-field ionization, electron correlations, giant dipole resonance}

\maketitle
\section*{Acronyms}
\begin{acronym}[ARPACK]
\setlength{\itemsep}{-1.25\parsep}
\acro{ADK}[ADK]{Ammosov, Delone, and Krainov}
\acro{APS}{American Physical Society}
\acro{ARPACK}{Arnoldi package}
 \acro{ATI}{Above-threshold ionization}
 \acro{CAP}{Complex absorbing potential}
 \acro{CI}{Configuration interaction}
 \acro{CIS}{Configuration interaction singles}
 \acro{ECS}{Exterior complex scaling}
 \acro{FEL}{Free-electron laser}
 \acro{FERMI}{Free-electron Laser for Multidisciplinary Investigations, Trieste}
 \acro{FLASH}{Free-electron Laser Hamburg}
 \acro{FWHM}{Full width at half maximum}
 \acro{GDR}{Giant dipole resonance}
 \acro{HFS}{Hartree-Fock-Slater}
 \acro{HHG}{High harmonic generation}
 \acro{IOP}{Institute of Physics}
 \acro{LAPACK}{Linear algebra package}
 \acro{LCLS}{(Stanford) LINAC Coherent Light Source}
 \acro{MBES}{Magnetic bottle electron spectrometer}
 \acro{OCT}{Optimal control theory}
 \acro{PAD}{Photoelectron angular distribution}
 \acro{PES}{Photoelectron spectrum}
 \acro{RPAE}{Random-phase approximation with exchange}
 \acro{SACLA}{Spring-8 \r{A}ngstr\"om Compact Free Electron Laser}
 \acro{SAE}{Single-active electron}
 \acro{SASE}{Self-amplified spontaneous emission}
 \acro{SES}{Smooth exterior scaling}
 \acro{SLAC}{Stanford Linear Accelerator Center}
 \acro{SPring-8}{Super Photon Ring 8 GeV}
 \acro{t-surff}{Time-dependent surface flux}
 \acro{TDCIS}{Time-dependent configuration interaction singles}
 \acro{UV}{Ultraviolet}
 \acro{VUV}{Vacuum Ultraviolet}
 \acro{XATOM}{Integrated toolkit for x-ray atomic physics}
 \acro{XCID}{Configuration interaction dynamics package for multichannel dynamics}
 \acro{XFEL}{X-ray free-electron laser}
 \acro{XUV}{Extreme ultraviolet}
\end{acronym}

\section{\label{sec:level1}Introduction}

Remarkable progress in the realm of light-source development has shaped the physics of the 20$^{\mathrm {th}}$ 
and the beginning of the 21$^{\mathrm {st}}$ century. The power of producing coherent electromagnetic radiation 
of high intensity evolved simultaneously with the quest for ever increasing precision and for more information about 
the structure and dynamics of matter. Today, modern light sources deliver intense, 
ultrashort pulses at frequencies ranging from the terahertz
to the hard x-ray regime. They provide the experimental means to control and image atomic and 
molecular systems and to test theoretical predictions of nonlinear processes 
\cite{corkumkrausz,chapman,levesque,spanner}. On the one hand light is utilized to investigate the 
structure and dynamics of atoms; on the other hand it is employed to control atomic degrees of freedom 
and to steer electrons, which in turn leads to the development of new technology, such as attosecond 
light sources \cite{Nature420,corkumkrausz,kra}. In the interaction with light the forces exerted on the electrons can be 
comparable to the intra-atomic forces and the ultrashort pulse durations reach 
the typical time scales involved in electronic excitations in atoms, molecules and clusters,
roughly between 50~attoseconds ($1~$as$~=10^{-18}$~s) 
and 50~femtoseconds ($1~$fs~$=10^{-15}$~s)~\cite{kra,wabnitz}. Therefore, the realm of strong-field physics, 
multiphoton processes and light-matter interactions on an ultrashort timescale, 
ranging from fs to as,
has become a focus of interest.  
All of these processes are directly linked to 
the process of photo-excitation and -ionization. 

Belonging to the last generation of light sources, free-electron lasers (\acs{FEL}s) provide extraordinarily intense light pulses that permit the investigation and control of inner-shell processes, Auger decay or above-threshold ionization 
(\acs{ATI}) \cite{kanter,auger1975,PhysRevLett.42.1127}. 
Typically, the photon energies range 
from the \acs{UV} to the x-ray range and the duration of \acs{FEL} pulses are as short 
as a few femtoseconds, i.e., suitable for studying phenomena in atoms and 
molecules with a new quality in time resolution. 

The present tutorial deals with the theoretical investigation of the nonlinear response of 
atomic systems interacting with intense light pulses, 
spanning a broad frequency range from the infrared to the x-ray regime. 
After an introduction to the theoretical and computational framework an analysis of 
the adiabatic eigenstates of the many-body Hamiltonian will be performed in order to 
clarify the notion of adiabaticity in strong-field ionization. 
Subsequently, the calculation of photoelectron spectra will be presented and applied to \acs{XUV} ionization of argon. Xenon will also be investigated in the \acs{XUV} energy regime, concentrating on the $4d$ giant dipole resonance (\acs{GDR}). Finally, the x-ray regime will be addressed in order to quantify the impact of multiphoton ionization in current and future \acs{FEL} experiments.

For the above-mentioned purposes, the theoretical framework must be capable of efficiently describing the absorption, emission and the scattering of photons by atoms. Consequently, a good description of the electronic structure, of the (strong)
light field and of the interaction of light with matter is needed. 
Moreover, in the present article aspects of many-body physics in the atomic shell shall be investigated.
Phenomena which involve collective electronic behavior cannot be described 
within the single-active electron picture. Correlation effects in the atomic shell 
can be scrutinized by employing the configuration-
interaction singles (\acs{CIS}) scheme. The wave function is expanded in the \acs{CIS} basis and the time-dependent Schr\"odinger equation is solved ``ab initio'' by numerical time propagation.
Cases where collective electronic behavior plays a crucial role will be exemplified for xenon. 

Throughout this tutorial atomic units will be used, which are the ``natural'' units for processes occurring on atomic time and length scales. In atomic units the mass and the charge of the electron are set to unity, i.e., $m_e=1$, and $|e|=1$. Furthermore, the reduced Planck constant is set to one, $\hbar=1$. It follows that the fine-structure constant $\alpha=\fr{e^2}{\hbar c}=\fr{1}{c}\approx \fr{1}{137}$, where $c$ is the speed of light. The Bohr radius, or bohr, is the atomic unit of length, $a_0=\fr{1}{\a}\fr{\hbar}{m_ec}=0.53$\,\r{A} and the Hartree energy, or hartree, is the atomic unit of energy, $E_h=\fr{\hbar^2}{m_e a_0^2}=27.21$\,eV. The atomic unit of time corresponds to the time the electron needs to travel the distance of a Bohr radius at the velocity $v_0=\fr{a_0E_h}{\hbar}$, such that $t_0=\fr{\hbar}{E_h}=\fr{1}{\a^2}\fr{\hbar}{m_ec^2}=24$\,as. Then, the classical orbital period of the
electron in the first Bohr orbit is given by $T=2\pi t_0\approx150$~as.

\section{Photoionization}\label{photoion}
As one of the most probable processes to happen when light interacts with matter, photoionization has been studied extensively from the very beginning of quantum theory \cite{bethe}. Both, theoretically and experimentally, the ionization of atoms and molecules has served as a tool to investigate the nature of the atomic shell and the molecular orbitals \cite{Sta82}. 
If the light fields interacting with matter are strong, e.g. at \acs{FEL}s, multiphoton processes play a significant role
~\cite{young,moshammer}. The removal of a deep inner-shell electron is followed by various processes depending on the atomic states and the photon energy \cite{drescher,fukuzawa}. 
 
There are two ionization regimes which can be classified according to the so-called Keldysh parameter $\gamma= \sqrt{I_p/(2 U_p)}$ \cite{iva}: The ``tunneling ionization'' \cite{kel} and the ``(perturbative) multiphoton ionization'' \cite{mai}. $I_p$ is the ionization potential and $ U_p=I/(4\omega^2)$ is the ponderomotive potential, 
which corresponds to the average energy of a free electron oscillating in the electric field. It involves the intensity of the light field, $I$, and the central frequency, $\omega$. In this way, the Keldysh parameter sets into proportion the frequency of the light field and the field amplitude, $\gamma=2\omega \sqrt{I_p/ (2 I)}$. For $\gamma\ll 1$ the ponderomotive potential is much larger than the ionization potential and
ionization is governed by tunneling ionization, while for $\gamma \gg 1$ the process is governed by multiphoton ionization. 
In the range of $\gamma\approx 1$ both effects compete with each other \cite{yam}.

\begin{figure}[!b]
 \centering
 \includegraphics[width=.75\linewidth]{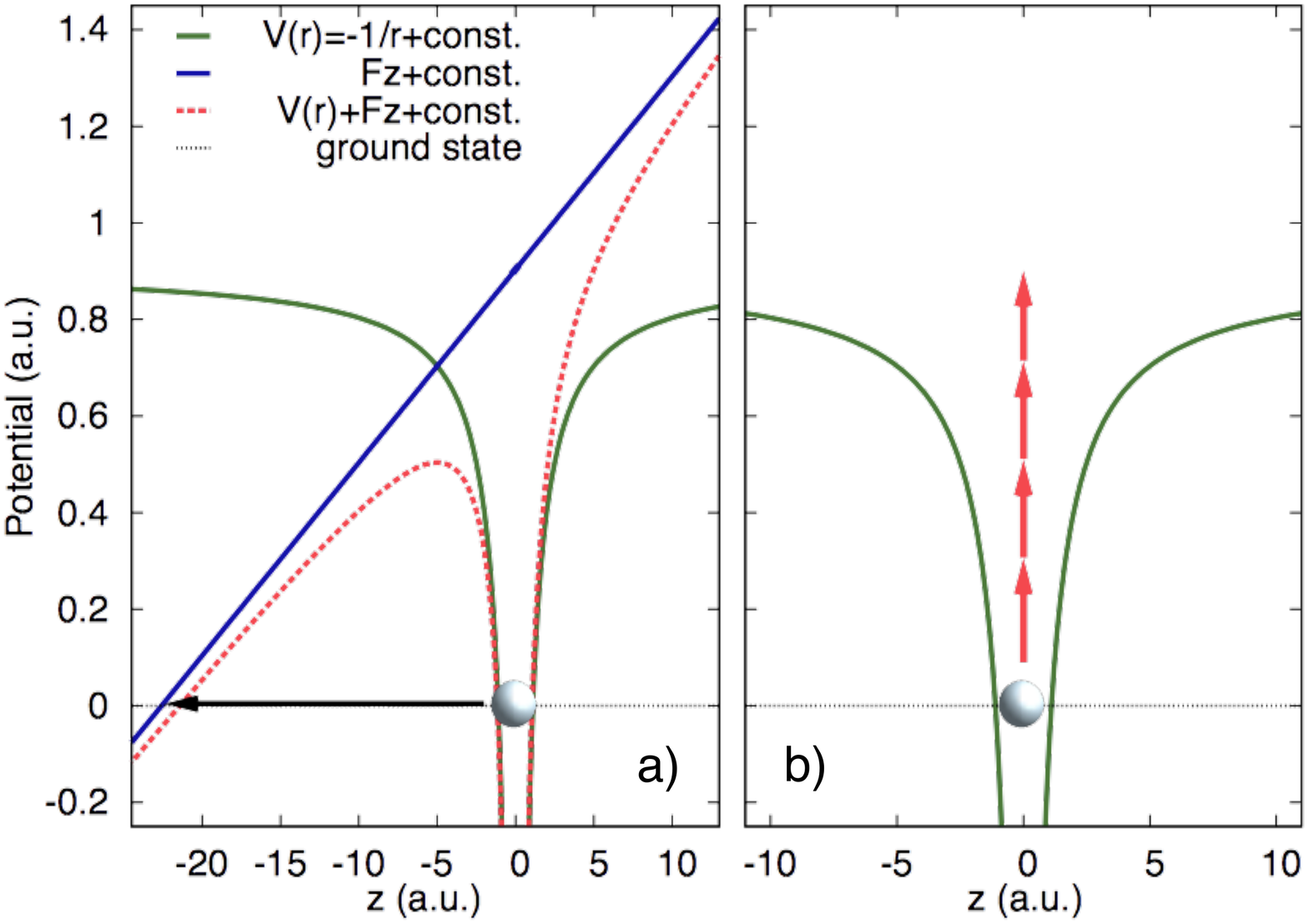} 
 \caption{Different ionization regimes depicted schematically for helium. a) Tunneling ionization, $\g\ll1$. The strong electric field (black-dashed line) tilts the Coulomb potential (green curve) and a barrier of finite width is generated (red curve). The electron can tunnel through the barrier and leave the ionic core in the direction of the barrier suppression. b) Multiphoton regime, $\g\gg 1$. By absorbing several photons the Coulomb barrier is overcome and the electron can be ionized.}
 \label{tunnelmulti}
\end{figure}

Regarding the theoretical description there is a significant difference between the two regimes: In general, the tunneling regime requires a nonperturbative description, because the magnitude of the light's electric field is comparable with the intra-atomic electric field and cannot be treated as a small perturbation. The electric field, which oscillates with a small frequency, bends the Coulomb potential in such a way that a barrier of finite width is created through which the electron can tunnel and leave the ion. This is shown schematically in Fig.~\ref{tunnelmulti}~a). Since this is a quasistatic picture, the Keldysh parameter has been connected to the notion of adiabaticity in the ionization process \cite{mev}. Far into the tunneling regime, the atomic response is considered to be purely adiabatic, which in this context means that the ionization rate at a given time is solely defined by the instantaneous electric field.

For relatively weak fields compared to the intra-atomic Coulomb potential (for atomic systems $I\ll 10^{16}$Wcm$^{-2}$) the light field can be treated as a perturbation and the light-matter interaction can be classified within the lowest orders of perturbation theory. It can be viewed as a process where the system absorbs simultaneously several photons to undergo ionization, see Fig.~\ref{tunnelmulti}~b). The order of the interaction or, equivalently, the number of photons that are absorbed is described by the corresponding order of perturbation theory.
In this regime cross sections for the absorption or scattering of photons by atoms can be defined \cite{saenz}. For this purpose, the transition matrix element $M_{F\leftarrow I}$ associated with the transition induced by the interaction Hamiltonian $\hat{H}_\mathrm{int}$ from an initial state of the coupled atom-light system $I$ to a final state $F$ is calculated. For instance, the absorption of two photons is described in second-order perturbation theory. All possible pathways leading to the final state by absorbing two photons have to be taken into account. Through the absorption of one photon an intermediate state $K$ is populated and a summation over all those possible intermediate states must be carried out
\begin{equation}\label{transmatrixelement}
M^{(2)}_{F\leftarrow I}=\sum_{K} \frac{ \langle F | \hat{H}_{\mathrm{int}} |K \rangle \langle K | \hat{H}_{\mathrm{int}}| I \rangle}{E-E_{K}+\frac{i}{2}\Gamma_{K}+E_I},\end{equation}
where $E$ is the energy of the photon, $E_{K}$ and $\Gamma_{K}$ are the energy and the decay rate of the intermediate state, respectively, and $E_I$ is the energy of the initial state. $\Gamma_K$ is associated to the state's life time $\t_K$ by the relation $\G_K=1/\t_K$. The two-photon cross section, $\sigma^{(2)}$, is proportional to $|M^{(2)}_{F\leftarrow I}|^2$.  

Exemplarily and for simplicity disregarding indirect channels the rate equation for the population evolving due to an $N$-photon ionization process reads 
\begin{equation}
\fr{\dd}{\dd t} P_0(t) =- P_N(t) = -\sigma^{(N)}j^N P_0(t),
\end{equation} 
which involves the ground state population $P_0$, the product of the cross section for the $N$-photon process, $\s^{(N)}$, and the $N^{\mathrm{th}}$ power of the photon flux, $j=I/\omega_{\mathrm {photon}}$, i.e., the number of incident photons per unit time and unit area. Usually, the $\s^{(N)}$ for $N>1$ are called ``generalized'' cross sections, since their units are not units of a cross section. The differential equation is solved through integration
\begin{equation}
P_0(t)=\exp\left(-\int_{-\infty}^t\dd \t  \sigma^{(N)}j^N\right) +\mathrm{const.}
\end{equation}
 It follows that as long as the saturation regime is not reached, i.e., as long as the intensity is low enough that the ground state is not depleted (i.e., $P_0\approx1$) and the cross section does not depend on the intensity, the following relation holds: 
 \begin{equation}
 \ln P_N=N\ln I + \ln\sigma^{(N)}+\mathrm{const.}\label{logdep}
 \end{equation} 
 The linear dependence results in a straight line on a double-logarithmic scale and allows to read off the order of the ionization process from the slope of the curve (i.e., one for one-photon processes, two for two-photon processes, and so forth).


\section{\label{theory}Theory and Method}
The general time dependent Schr\"odinger equation of an $N$-electron system is given by 
\begin{equation}
 i \fr{\pa}{\pa t} |\Y^N(t)\rangle = \hat {H} (t) |\Y^N(t)\rangle , \label{schr}
\end{equation}
where $|\Y^N(t)\rangle$ is the $N$-electron wave function. Here, the Hamiltonian has the form
\begin{align}\label{ham}
 \hat{H}(t)=\underbrace{\sum_{n=1}^N \left(\fr{\hat{\mathbf{p}}_n^2}{2}-\fr{Z}{|\hat{\mathbf{r}}_n|} +\hat{V}_{\mathrm{ MF}}(\hat{\mathbf{r}}_n)\right)}_{\hat{H}_0} + \underbrace{\fr{1}{2} \sum_{n\neq n'}\fr{1}{|\hat{\mathbf{r}}_n-\hat{\mathbf{r}}_{n'}|}-\sum_{n=1}^N\hat{V}_{\mathrm{ MF}}(\hat{\mathbf{r}}_n)}_{\hat{H}_1}+\underbrace{\vphantom{\sum_{i\neq j}^j} \hat{\mathbf{p}} \cdot {\mathbf{A}}(t)}_{\hat{H}_\mathrm{int}},
\end{align}
where the single-electron part of the Hamiltonian, in the following denoted by $\hat{H}_0$, contains the kinetic energy $\hat{T}\equiv\sum_n \hat{\mathbf{p}}_n^2/2$, the nuclear potential $\hat{V}_{\mathrm{ nuc}}\equiv-\sum_n Z /|\hat{\mathbf{r}}_n|$ and the potential at the mean-field level $\hat{V}_{\mathrm{ MF}}$. The electron-electron Coulomb interaction $\hat{V}_{\mathrm {e-e}}\equiv\fr{1}{2} \sum_{n\neq n'}1/|\hat{\mathbf{r}}_n-\hat{\mathbf{r}}_{n'}|$ completes the many-electron part of the Hamiltonian $\hat{H}_{1}\equiv\hat{V}_{\mathrm {e-e}}-\hat{V}_{\mathrm{ MF}}$. Finally, $\hat{H}_\mathrm{int}$ denotes the light-matter interaction in the minimal coupling scheme involving the momentum operator $\hat{\mathbf{p}}=\sum_n \hat{\mathbf{p}}_n$ and the vector potential ${\mathbf{A}}(t)$. The Coulomb gauge for the vector potential, $\mathrm{div} \mathbf{A}\!=\!0$, is often chosen in the context of atomic physics. As a consequence, the vector potential is divergence-free and purely transverse. Here, the light is described semiclassically and as a function of time only. Due to the exclusive time dependence the electric field does not change over the spatial extent of the atom and, consequently, our description remains in the dipole approximation.

\subsection{Time-dependent \acs{CI} singles (TDCIS) scheme}

\begin{figure}[!tb]
 \centering
 \includegraphics[width=.75\linewidth]{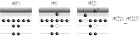}
 \caption{Schematic atomic level scheme. Full spheres symbolize electrons and open circles holes; the lines denote energy levels and the grey shaded area symbolizes the electronic continuum. Starting from the Hartree-Fock ground state $|\Phi_0\rangle$ through repeated one-particle--one-hole excitations the full \acs{CI} space is built.}
 \label{CIscheme}
\end{figure}

In order to solve the electronic structure problem a convenient basis set must be found in which the wave function can be expanded. In quantum chemistry a widely-used scheme is the configuration interaction (\acs{CI}) \cite{szabo}: Starting from the Hartree-Fock ground state $|\Phi^{\mathrm{HF}}_0\rangle$ the configuration space is built up by exciting electrons from the occupied orbital and promoting them to an orbital which was previously unoccupied (also called a virtual orbital). This yields a one-particle--one-hole state $|\Phi_i^a\rangle$ for the $N$-electron system. Exciting a second electron and letting it occupy another previously unoccupied orbital yields a two-particle--two-hole state $|\Phi_{i,j}^{a,b}\rangle$, promoting a third electron from an initially occupied orbital to a virtual orbital represents a three-particle--three-hole state $|\Phi_{i,j,k}^{a,b,c}\rangle$, and so forth. This is visualized schematically in Fig.~\ref{CIscheme}, where $i,j,k,\ldots$ denote an initially occupied orbital and $a,b,c,\ldots$ denote a virtual orbital.
The configuration space is complete if all possible particle--hole excitations are taken into account. However, for higher excitations the space of virtual orbitals grows immensely. Therefore, neglecting higher order excitations $|\Phi_{i,j}^{a,b}\rangle, |\Phi_{i,j,k}^{a,b,c}\rangle,\ldots$, we truncate the configuration space and consider only states where one, single electron is excited to a previously unoccupied state. This space is called configuration interaction singles (\acs{CIS}) space. 

To construct it we start from the Hartree-Fock ground state $|\Phi_0^N\rangle$ of a closed-shell $N$-electron system as a reference state which is obtained from the vacuum state $|0\rangle$ by applying electron creation operators.
The anticommuting operators $\hat{c}_{p\s}^\dagger$ and $\hat{c}_{p\s}$ are spin-orbital creation and annihilation operators, which create an electron in the spin orbital $|\varphi _{p\s}\rangle$ or annihilate an electron from this orbital, respectively, i.e., $\hat{c}^\dagger_{p\sigma} |0\rangle =|\phi_{p\sigma}\rangle$. The field-free one-particle Hamiltonian $\hat{H}_0$ has the form
$\hat{H}_0 = \sum_p \varepsilon_p \sum_\sigma  \hat{c}^\dagger_{p \sigma}\hat{c}_{p \sigma}$,
such that $\hat{H}_0|\varphi_{p \sigma}\rangle = \varepsilon_p |\varphi_{p \sigma}\rangle$, where $\varepsilon_p$ denotes the energy of the orbital $|\varphi_{p\sigma}\rangle$. We consider closed-shell atoms and processes in which the total spin of the system is not altered ($S=0$, and there is no magnetic field involved), such that only spin singlets occur. Therefore, we henceforth drop the spin index and treat only the spatial part of the orbitals $|\varphi _p\rangle$. 
 The Hartree-Fock ground state in our notation is the antisymmetrized product, or Slater determinant
\begin{equation}
|\Phi_0^N\rangle = \det\left( |\varphi _{1}\rangle,\ldots ,|\varphi_{N}\rangle \right),
\end{equation}
 which involves the $N$ energetically lowest spin orbitals.  
 In order to account for electronic excitations within the \acs{CIS} scheme we build the so-called one-particle--one-hole configurations which have the form
\begin{equation}
|\Phi_i^a\rangle=\hat{c}_{a}^\dagger \hat{c}_{i} |\Phi_0\rangle.
\end{equation}
The index $i$ symbolizes an initially occupied orbital and $a$ denotes a virtual orbital in the sense described previously. 
The total N-electron wave function (now omitting the superscript $N$ for better legibility) for solving Eq.~\eqref{schr} is expanded within the \acs{CIS} approach as~\cite{pab}
\begin{align}
 |\Psi(t)\rangle&=\alpha_0(t)|\Phi_0\rangle+\sum_{i,a}\alpha_i^a(t)|\Phi_i^a\rangle.\label{tdcis}
\end{align}

Inserting this expansion into the Schr\"odinger equation (\ref{schr}) and projecting it onto the states $|\Phi_0\rangle$ and $|\Phi_i^a\rangle$ yields the following equations of motion for the time-dependent expansion coefficients $\a_i^a(t)$~\cite{pab}:
\begin{subequations}\label{ciseqs}
\begin{align}
 i\dot{\a}_0(t) &= \mathbf{A}(t)\cdot\sum_{i,a}\langle \Phi_0 | \, \hat{\mathbf{p}}\, | \Phi_i^a\rangle \a_i^a(t),\label{cis1}\\
 i\dot{\a}_i^a(t) &= (\varepsilon_a-\varepsilon_i) \a_i^a(t) + \sum_{j,b} \langle \Phi_i^a |\hat{H}_1|\Phi_j^b    \rangle    \a_j^b(t) \nn\\
                 &  +\mathbf{A}(t)\cdot\left( \langle\Phi_i^a | \, \hat{\mathbf{p}}\, |\Phi_0  \rangle \a_0(t) + \sum_{j,b}\langle \Phi_i^a |\, \hat{\mathbf{p}}\,|\Phi_j^b    \rangle    \a_j^b(t) \right ).\label{cis2}
\end{align}
\end{subequations}
 The computational challenge lies in the numerical evaluation of the Coulomb matrix elements $\langle \Phi_i^a |\hat{H}_1|\Phi_j^b\rangle$, which must be calculated for all active occupied and virtual orbitals. In order to determine all the virtual orbitals that must be included in the calculation, a cut-off energy $E_{\mathrm{cut}}$ is defined up to which the virtual orbitals are calculated. Of course, it is essential to make the virtual space large enough to contain all states of interest. 

In the following we assume that the electric field is linearly polarized along the $z$-axis, such that the light-atom interaction term simplifies to the projection of the momentum on the direction of the vector potential, $\hat{p}_zA_z$. Using the Slater-Condon rules \cite{szabo} and writing the one- and two-body matrix elements explicitly in terms of the spatial orbitals the Eqs.~\eqref{ciseqs} read \cite{roh}:
\begin{subequations}\label{spatialeom}
\begin{align}
i\dot{\alpha}_0 &= 2A(t)\sum_{i,a} \alpha_i^a\,p_{ia},\label{cis3}\\
 i \dot{\alpha}_i^a &= (\varepsilon_a-\varepsilon_i)\alpha_i^a + \sum_{i'b}\alpha_{i'}^b(2v_{ai'ib}-v_{ai'bi})+\sqrt{2}A(t)\,\alpha_0\, p_{ai}+A(t)\sum_b p_{ab}\,\alpha_{i}^b\nn \\
&-A(t)\sum_{i'}p_{i'i}\,\alpha_{i'}^a,\label{cis4}
\end{align}
\end{subequations}
where the two-body matrix elements are given by 
\begin{equation}
v_{pqrs}=\int \dd^3 r\,\dd^ 3 r'\,\varphi_p^\dagger (\mathbf{r})\varphi_q^\dagger (\mathbf{r'})\fr{1}{|\mathbf{r}-\mathbf{r'}|}\varphi_r (\mathbf{r})\varphi_s (\mathbf{r'}),
\end{equation}
and the matrix elements of the dipole operator, which is a one-body operator, are of the form
\begin{equation}
p_{ab}=\langle \varphi_a|\hat{p}_z  | \varphi_b \rangle.
\end{equation}
The differential equations~\eqref{spatialeom} are solved by numerical time propagation either using the $4^\mathrm{th}$ order Runge-Kutta algorithm or the Lanczos propagation, which will be introduced in the next section. Since the light-matter interaction is included in a nonperturbative manner in the Hamiltonian and, consequently, in Eq.~\eqref{schr} all orders of the interaction are automatically included in the calculations. In principle, this approach can cover all frequency ranges as long as the dipole approximation can be assumed to be valid.

\subsection{Multichannel physics and electron correlations}\label{multichannel}

Collective effects or correlation phenomena are abundant in atomic and molecular systems. Examples are autoionization, Auger decay, Fano resonances, the giant dipole resonance in xenon, and interatomic Coulombic decay \cite{shirleybecker,PhysRev.124.1866,Fet71,PhysRevLett.79.4778}. All of these involve the description of electronic dynamics beyond the single-active electron approach. 

\begin{figure}[!tb]
 \centering
 \includegraphics[width=.5\linewidth]{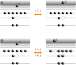}
 \caption{Schematic of the intrachannel (a) and the interchannel (b) coupling. The operator $\hat{V}_{\mathrm {e-e}}$ couples different electronic configurations. Intrachannel coupling: The particle-hole interaction is confined to states with the same hole index. Interchannel coupling: Couples different configurations for all pairs of hole indices and virtual state indices.}
 \label{intrainter}
\end{figure}

The \acs{CIS} approach encapsulates many-body interactions through the coupling of different configurations contained in the Coulomb matrix elements $\langle \Phi_i^a | \hat{V}_{\mathrm{e-e}} |\Phi_j^b\rangle$. The set of active occupied orbitals $i$ builds the space of channels through which excitation and ionization can occur which automatically allows for the distinction and analysis of multichannel physics. Through the coefficients $\a_i^a(t)$ ionization and excitation processes can be analyzed in a channel-resolved manner. In this way, channel-resolved quantities can be inferred from the $N$-electron wave function, such as ionization probabilities, cross sections, and the ion density matrix (see, e.g., Ref.~\cite{decoherence}).

In theory, we can manually switch off the correlation effects. If we allow only for the Coulomb matrix elements with hole indices $i=j$ to be nonzero, $\langle \Phi_i^a | \hat{V}_{\mathrm{e-e}} |\Phi_i^b\rangle$, we obtain the ``intrachannel'' picture, which is visualized in Fig.~\ref{intrainter}\! a). Physically, this means that the excited electron, although interacting with the hole, cannot modify the ionic hole state (depicted by open circles) in the remaining parent ion. 

Without this restriction there can be non-vanishing matrix elements $\langle \Phi_i^a | \hat{V}_{\mathrm{e-e}} |\Phi_j^b\rangle$ for two configurations whose hole indices are different from one another, additionally to a different index of the excited electron, $i=j$ {\em{and}} $i\neq j$. This is called ``interchannel'' coupling and means that a simultaneous change of the excited electron state and the ionic hole state through electron-electron correlation is permitted. Fig.~\ref{intrainter}~\!b) depicts this situation, where states with $i\neq j$ and $a\neq b$, are coupled. In this way a correlated particle-hole pair is created. 
Whenever it is not possible to write the wave function as a single particle--hole state, yet a superposition of particle-hole states is required to describe the state of the system, the system is characterized by collectivity.
We will refer to the electron interaction and the entangled states to which it leads as {\em{electron correlation effects}}.
The comparison of the results obtained within the intrachannel and interchannel schemes enables us to distinguish the impact of the configuration interaction, i.e., the electron correlation effects, onto a certain process. 

Furthermore, mean-field approaches can be used to simplify the dynamics beyond the intrachannel model. For instance, the Hartree-Fock-Slater (\acs{HFS}) model \cite{PhysRev.81.385} reduces the complexity because the exchange interaction between the electrons is modeled in the local density approximation. This results in an effective one-particle picture where the electron experiences a mean-field potential $V(\mathbf{r})$ created by the other electrons in the shell: 
\begin{equation}
V(\mathbf{r})=-\fr{Z}{r} + \int\dd ^3r'\fr{\rho(\mathbf{r})}{|\mathbf{r}-\mathbf{r}'|} +V_{\mathrm{ex}}(\mathbf{r}),
\end{equation}
where $V_{\mathrm{ex}}(\mathbf{r})=-\fr{3}{2}\left[ \fr{3}{\p} \rho(\mathbf{r} )\right]^{1/3}$ is the Slater exchange potential.
Here $Z$ is the nuclear charge, and $\rho$ is the electron density. For the case of large distances from the origin the Latter tail correction \cite{PhysRev.99.510} can be used to obtain the proper asymptotic potential for both occupied and unoccupied orbitals:
\[V (\mathbf{r}) = - (Z' +1)/r, \text{ if } -Z/r + \int\dd ^3r'\fr{\rho(\mathbf{r})}{|\mathbf{r}-\mathbf{r}'|} + V_{\mathrm{ex}}(\mathbf{r})< - (Z' + 1)/r,\] 
with the effective charge $Z' = Z - N_{\mathrm{elec}}$, $N_{\mathrm {elec}}$ being the number of electrons. In this way, we obtain the asymptotic behavior of $Z/r$ for small radii from the ion because the electron can feel the whole nucleus as a Coulomb attractor. Another asymptote, $1/r$, is obtained for large distances, since from afar the $N_{\mathrm{elec}}-1$ positive nuclear charges are screened by the other electrons, such that the electron experiences an effective Coulomb potential equivalent to one unscreened charge. Naturally, this model can only be applied in a meaningful way if electron correlations are expected to be negligible.


\subsection{\label{sec.3} Algorithm for wave-function propagation and for the diagonalization of the Hamiltonian}
The Schr\"odinger equation~\eqref{schr} can be solved numerically by time propagation. Different schemes have been employed for propagating the wave function \cite{Feit1982412,talezerkosloff}, but we will focus here on the Lanczos-Arnoldi algorithm \cite{golub}. The well-known fourth-order Runge-Kutta algorithm \cite{hoffmannnum}, which is often employed for solving differential equations of first order, is not symmetric in time, but stable if the time step is chosen to be sufficiently small. Since in standard atomic physics problems forward propagation of the Schr\"odinger equation suffices to obtain the relevant observables and dynamics (e.g. ionization yields, cross sections, electron spectra, etc.) time symmetry is not crucial. If, however, in addition backward propagation is involved in the computations higher accuracy in propagation is needed \cite{Goe15}. To this end, the Lanczos-Arnoldi algorithm, which is a Krylov subspace method \cite{golub}, can be applied. It is an iterative approximation scheme for eigenvectors and eigenvalues of a matrix, which is, in general, the method of choice for large matrices. It has been used for atomic and molecular physics problems before \cite{kosloff94}. 

\subsubsection{Lanczos-Arnoldi algorithm}

Suppose that the eigenvalues of a large and sparse Hermitian matrix $A\in \mathbb{C}^{n\times n}$ shall be calculated. Let $\lambda_1, \ldots, \lambda_n$ be the $n$ eigenvalues of $A$, ordered by their magnitude. The Rayleigh quotient 
\begin{equation}
R(\mathbf{x})=\fr{\mathbf{x}^*A\mathbf{x}}{\mathbf{x}^*\mathbf{x}},
\end{equation}
with $\mathbf{x}\in \mathbb{C}^n$, yields the smallest and the largest eigenvalues by the relations $\lambda_1=\underset{\mathbf{x}\neq \mathbf{0}}{\max}\, R(\mathbf{x})$ and $\lambda_n=\underset{\mathbf{x}\neq \mathbf{0}}{\min}\, R(\mathbf{x})$ \cite{golub}. 
Let $\mathcal{V}_k$ be a subspace of $\mathbb{C}^{n\times n}$ and let $\{{\mathbf{q}}\}_1^{k}=\left\{ {\mathbf{q}}_{1}, {\mathbf {q}}_{2}, \ldots, {\mathbf {q}}_k \right\}$
be an orthonormal basis of $\mathcal{V}_k$. Arranging the vectors ${\mathbf{q}}_k$ as columns in a matrix $Q_k$ the eigenvalues of $Q_k^{T}AQ_k$
shall approximate the eigenvalues of $A$. 
The Lanczos-Arnoldi method generates the vectors ${\mathbf{q}}_k$ iteratively, such that the eigenvalues of the matrices $Q_k^{T}AQ_k=T_k\in \mathbb{C}^{k\times k}$, with $k<n$, are, with $k\rightarrow k+1$, progressively better approximations to the eigenvalues of A. 
Based on the Courant-Fischer theorem~\cite{golub} the algorithm determines increasingly better eigenvalues consider the Rayleigh quotient of the matrix $T_k$:
\begin{subequations}
\begin{alignat}{3}
\lambda_1 &\ge \underset{\mathbf{y}\neq \mathbf{0}}{\max}\, R(Q_k\mathbf{ y})&\,=R(\mathbf{u}_k) &\equiv M_k ,\label{maximum}\\
\lambda_n &\le\underset{\mathbf{y}\neq \mathbf{0}}{\min}\, R(Q_k\mathbf{ y})&\,= R(\mathbf{v}_k) &\equiv m_k  ,
\end{alignat}
\end{subequations}
with $\mathbf{y}\in \mathbb{C}^n$ and vectors $\mathbf{u}_k,\mathbf{v}_k\in\mathrm{span}\left\{{\mathbf{q}}_{1}, {\mathbf {q}}_{2}, \ldots, {\mathbf {q}}_k  \right\}$. $R(\mathbf{x})$ increases most rapidly in the direction of the gradient
\begin{equation}
 \mathbf{\nabla}R(\mathbf{x}) = \fr{2}{\mathbf{x}^*\mathbf{x}} \left[ A\mathbf{x}-R(\mathbf{x}) \mathbf{x}\right],
\end{equation}
from which it follows that $\mathbf{\nabla}R(\mathbf{x})\in\mathrm{span}\left\{\mathbf{x},A\mathbf{x}\right\}$. The largest eigenvalue of the next iteration step $ M_{k+1}$ will be larger than $M_k$, and therefore approach the ``real'' eigenvalue of the original matrix $A$ if the next vector $\mathbf{q}_{k+1}$ is determined such that $\mathbf{\nabla}R(\mathbf{x}) \in\mathrm{span}\left\{{\mathbf{q}}_{1}, {\mathbf {q}}_{2}, \ldots, {\mathbf {q}}_{k+1}  \right\}$. Following the same argument for the eigenvalue of minimal magnitude, if also $\mathbf{\nabla}R(\mathbf{v}_k) \in\mathrm{span}\left\{{\mathbf{q}}^{1}, {\mathbf {q}}^{2}, \ldots, {\mathbf {q}}^{k+1}\right\} $ then $m_{k+1}<m_k$, because $R(\mathbf{x})$ decreases most rapidly in the direction of the negative gradient $- \mathbf{\nabla}R(\mathbf{x}) $. Therefore, both requirements can be satisfied if $\mathbf{q}_{k+1}$ is chosen such that 
\begin{equation}
\mathrm{span}\left\{\mathbf{q}_1,\ldots,\mathbf{q}_{k+1}\right\}=\mathrm{span}\left\{\mathbf{q}_1,A\mathbf{q}_1,A^2\mathbf{q}_1,\ldots,A^{k}\mathbf{q}_1\right\},
\end{equation}
and, thereby, successively applying higher powers of the matrix $A$ the Krylov space is built.

In our case we wish to solve the time-dependent Schr\"odinger equation~\eqref{schr}. Formally this equation has the solution
\begin{equation}
 |\Psi(t)\rangle = \hat{U}(t,0)| \Psi(0)\rangle,
\end{equation}
where 
\begin{equation}
\hat{U}(t,0)=\mathcal{T}\exp \left[ -i \int_0^t \dd \t \hat{H}(\t)\right]
\end{equation}
 is the time evolution operator and $\mathcal{T}$ denotes the time-ordering operator
\begin{equation}
\mathcal{T}[\hat{H}(t_1)\hat{H}(t_2)]= \begin{cases} \hat{H}(t_1)\hat{H}(t_2) &\mbox{if } t_1<t_2 \\ 
\hat{H}(t_2)\hat{H}(t_1) & \mbox{if } t_1>t_2. \end{cases}
\end{equation}
Approximating the time evolution operator for small time arguments the wave function at the next time step $t+\dd t$ has the value
\begin{equation}
|\Psi(t+\dd t)\rangle= \eul^{-i\hat{H}\left(t+\fr{\dd t}{2}\right) \dd t} |\Psi(t) \rangle+\mathcal{O}(\dd t^3).
\end{equation}
The Krylov space is built by acting on the starting vector $|\Psi(0)\rangle$ with increasingly higher powers of the Hamiltonian 
\begin{equation}
\hat{H} ^n| \Psi(0)\rangle = {\mathbf v}^{n},
\end{equation}
where the set of $N$ vectors ${\mathbf{v}}_{n}$, $n=0,1,\ldots N-1$, forms the Krylov basis. $N$ is the dimension of the Krylov space which for numerical reasons should be as small as possible. Now, suppose we had a tridiagonal matrix $T=Q^T\hat{H}Q$ 
with $Q$ being orthogonal, then we could find an orthogonal matrix $U$ ($U^TU=\mathbb{1}$) that diagonalizes $T$. Let us call this diagonal matrix $D=U^T T U$. Then from $D= U^T Q^T \hat{H} QU$ it follows that
\begin{align}
Q^T\hat{H}Q&=UDU^T,
\end{align}
and from the properties of diagonal matrices we have
\begin{align}
 U\eul^{-iD \dd t}U^T&=\eul^{-iQ^T\hat{H}Q\dd t}\Rightarrow\\
 \eul^{-i\hat{H} \dd t}|\Psi\rangle&\approx QU\eul^{-iD \dd t}U^T Q^T |\Psi\rangle.\label{approxwf}
 \end{align}
Orthonormalizing the $N$ vectors $ \{{\mathbf v}_n\}$ we obtain an orthonormal set of vectors $\{\mathbf{q}_n\}$. If we arrange them as columns in a matrix $Q$ we know that the matrix $Q^T\hat{H}Q=T$ is tridiagonal (QR factorization) \cite{golub}:
\begin{equation}
T=\left(
\begin{tikzpicture}[baseline=(current bounding box.center)]
\matrix (m) [matrix of math nodes,nodes in empty cells ]{
 \alpha_0  & \ \beta_1       &                  &                  &    \\
  \beta_1   & \ \alpha_1   & \ \ \ \beta_2    &                     & {\text{\Huge 0} }\\
                  & \ \beta_2      &\ \ \ \alpha_2&\beta_3             &\\
                  &         &        &         &   \\
                  &                    &                &                   &   \beta_{N-1}   \\
    {\text{\Huge 0}}         &               &                    & \beta_{N-1}  & \alpha_N \\  };
\draw[loosely dotted] (m-3-3)-- (m-6-5);
\draw[loosely dotted] (m-3-2)-- (m-6-4);
\draw[loosely dotted] (m-3-4)-- (m-5-5);
\end{tikzpicture}
\right).
\end{equation}
We wish to directly compute the elements of this tridiagonal matrix, $\{\alpha_k\}_{k=1}^{N}$, $\{\beta_k\}_{k=1}^{N-1}$, in an iterative way. Since $\hat{H}Q=QT$ we find
\begin{equation}
 \hat{H}\mathbf{q}_k = \beta_{k-1}\mathbf{q}_{k-1} +\alpha_k\mathbf{q}_k+\beta_k\mathbf{q}_{k+1},
\end{equation}
while $\beta_0{\mathbf{q}}_0\equiv 0$. Solving this equation for ${\mathbf{q}}_{k+1}$, if 
\[\mathbf{r}_k=\left(\hat{H}-\alpha_k \mathbb{1} \right)-\beta_{k-1}{\mathbf{q}}_{k-1}\neq0,\]
 then ${\mathbf{q}}_{k+1}={\mathbf{r}}_k/\beta_k$, where $\beta_k=|{\mathbf{r}}_k|$. The vectors ${\mathbf{q}}_k$ are called Lanczos vectors. In this way the Lanczos-Arnoldi algorithm is obtained, where $\mathbf{q}_0$ denotes the starting vector:
\begin{equation}
\begin{aligned}\label{laalgorithm}
&k=0; \ \mathbf{r}_k= \mathbf{q}_0/|\mathbf{q}_0|;\ \beta_0=1;\ \mathbf{r}_1= \hat{H}\mathbf{q}_0;\ Q(\, :\,,1)=\mathbf{r}_1\\\
&{\mathrm{do\ while}}\ (\beta_k\neq0)\text{\footnotemark}\\
&\hspace{.8cm}k=k+1\\
&\hspace{.8cm}\mathbf{v}_k=\mathbf{q}_{k-1};\ \mathbf{q}_k=\mathbf{r}_{k-1}/\beta_{k-1};\ Q(\,:\,,k)=\mathbf{q}_k;\\
&\hspace{.8cm}\mathbf{r}_k=\hat{H}\mathbf{q}_k;\ \alpha_k=\mathbf{q}_k\cdot\mathbf{r}_k\\
&\hspace{.8cm}\mathbf{r}_k=\mathbf{r}_k-\beta_{k-1}\mathbf{v}_k-\alpha_k\mathbf{q}_k; \  \beta_k=|\mathbf{r}_k|\\
&{\mathrm{end\ do}}
\end{aligned}
\end{equation}
\footnotetext{Numerically, this is solved by a cut-off parameter: For instance, $|\beta_k|>10^{-18}$.}
Having calculated the sets of $\a$ and $\b$ coefficients, the tridiagonal matrix $T$ is built up and diagonalized to yield the eigenvalues that form the matrix $D$. Then the matrix product $P\equiv QU$ is calculated. Hence, we can evaluate Eq.~\eqref{approxwf}, the advantage being that we have just to perform matrix-vector multiplications in the small $N$-dimensional Krylov space (typically on the order of $N=10 \sim 20$) instead of the whole Hilbert space. 

The Lanczos-Arnoldi propagation method yields the same results as the Runge-Kutta algorithm with the additional feature of a significantly higher precision when used in the backward direction \cite{Goe15}. This algorithm can also be used to calculate approximate eigenstates of the Hamiltonian, which will be exploited in the next section.

\subsubsection{Diagonalization of the Hamiltonian: Diabaticity in tunneling ionization} \label{arpack}

Hamiltonian eigenvalue problems can involve very large matrices. For example, when high energies are involved 
or strong fields are considered, the necessary number of virtual states can be on the order of $10^3$ and the 
maximum angular momentum can be on the order of $10^2$. In combination with many active channels the 
dimension of the Hilbert space can easily reach $10^6$. Typically, these matrices are sparse, but nevertheless 
an efficient diagonalization algorithm is indispensable.
Since the diagonalization of large matrices is one of the most common procedures that are treated numerically 
it is not surprising that there is software providing a method for the solution of huge eigenvalue problems. 
The Arnoldi software package \acs{ARPACK} \cite{arp} employs the iterative Lanczos-Arnoldi algorithm to solve a 
general eigenvalue problem \[A\mathbf{x} = \lambda B \mathbf{x} .\] 

As an application for the diagonalization of the $N$-electron Hamiltonian we will examine the situation when helium is irradiated 
by an intense electric field $F(t)$ with a small photon energy compared to the electron binding energy. 
Let the process be characterized by a Keldysh parameter $\gamma\ll1$, i.e., it is described by tunneling ionization. 
As mentioned above, for small frequencies the tunneling regime can be viewed as quasi-static, 
which means, that the system is given time to adjust to the parameters on which it depends 
(in this case the electric field strength). Such a response is called adiabatic.
Therefore, let us analyze the associated adiabatic atomic eigenstates of helium along the lines of Ref.~\cite{antonia1}.
We would like to compute the eigenstates and the eigenvalues of the Hamiltonian:
\begin{align}
\left[\, \hat{H} + F(t)\,\hat{z} \,\right] |\Psi _n(t)\rangle=E_n (t)\,|\Psi _n(t)\rangle,\label{adequation}
\end{align}
where $E_n$ are the eigenenergies corresponding 
to the eigenvector $|\Psi_n\rangle$ of the system. Here, $\hat{H}=\hat{H}_0+\hat{H}_1 $ is the field-free Hamiltonian of Eq.~\eqref{ham}, $F(t)$ is the electric field in the $z$ direction and $\hat{z}$ is the position operator. 
The term $F(t)\,\hat{z}$ describes the dipole interaction between field and electron for a linearly polarized light field in the $z$-direction and thereby represents the light-matter interaction in the length form. The Hamiltonian is equivalent to Eq.~\eqref{ham} 
as long as no higher multipole orders play a role\footnote{Remember, that the vector potential $\mathbf{A}(t)$ and the electric field $\mathbf{F}(t)$ are related: $\mathbf{A}(t)=-\int_{-\infty}^t \mathbf{F}(\t)d\t$, $ \mathbf{F}(t)=-\fr{d}{d t} \mathbf{A}(t).$} \cite{cra}. 
Instead of solving the corresponding time-dependent Schr\"odinger equation
\begin{equation}
 i\partial_t | \Psi (t)\rangle =  \left\{ \hat{H} + F(t)\hat{z}\right\} | \Psi(t) \rangle, \label{tdse}
\end{equation}
the instantaneous, or adiabatic eigenvalue problem can be solved, where for each time $t$ the field $F$ has a certain value. At a given time $t$, the instantaneous eigenstates, which constitute the adiabatic basis, are defined by Eq.~\eqref{adequation}.
To distinguish adiabaticity from the onset of nonadiabatic effects we expand the electronic wave function in the adiabatic basis, $ | \Psi (t)\rangle = \sum_n \alpha_n(t)  |\Psi _n(t)\rangle$. 
Upon inserting this expression into Eq.~(\ref{tdse}) and projecting onto the eigenstate 
$|\Psi_m(t)\rangle$, the equation of motion for the coefficient $\alpha_m(t)$ reads
\begin{equation}
i \dot {\alpha} _m(t)+i \sum_n\alpha_n(t) \langle \Psi_m(t)|\partial_t|\Psi_n(t)\rangle
=\alpha_m(t) E_m(t). \label{eom}
\end{equation}
The off-diagonal matrix elements (also called nonadiabatic coupling terms) $\langle \Psi_m(t)|\dot{\Psi}_n(t)\rangle$ introduce couplings between different adiabatic eigenstates and lead to nonadiabatic dynamics~\cite{zen}. 
If these couplings are very small the so-called adiabatic approximation can be made and Eq. (\ref{eom}) becomes
\begin{equation}
i \dot {\alpha} _m(t)+i \alpha_m(t) \langle \Psi_m(t)|\dot{\Psi}_m(t)\rangle=\alpha_m(t) E_m(t). \label{eomadiab}
\end{equation}
 This equation shows that in the adiabatic approximation the system remains in a specific adiabatic eigenstate and evolves only with a phase. 

Now, it comes handy that for every tunneling state, 
i.e., every adiabatic state that allows the electron to tunnel 
through the field-induced barrier, there exists a discrete eigenstate of the instantaneous 
Hamiltonian (a so-called Siegert state) \cite{boh,sie}. 
Such a state is associated with a complex energy and lies by definition outside of the Hermitian domain of the Hamiltonian because the associated wave function is exponentially divergent for large distances from the atom. 
One strategy to eliminate the divergent behavior and render the tunneling wave function square integrable is to add a complex absorbing potential (\acs{CAP}) to the Hamiltonian, which yields complex eigenenergies~\cite{rismey,cedersantra}.
The imaginary part of the Siegert energy $E$ provides the tunneling rate 
$\Gamma$ of each Siegert state by the relation $\Gamma = -2\ {\mathrm{ Im}} (E)$ \cite{nim}.

\begin{figure}[!tb]
\centering
\includegraphics[width=.5\linewidth]{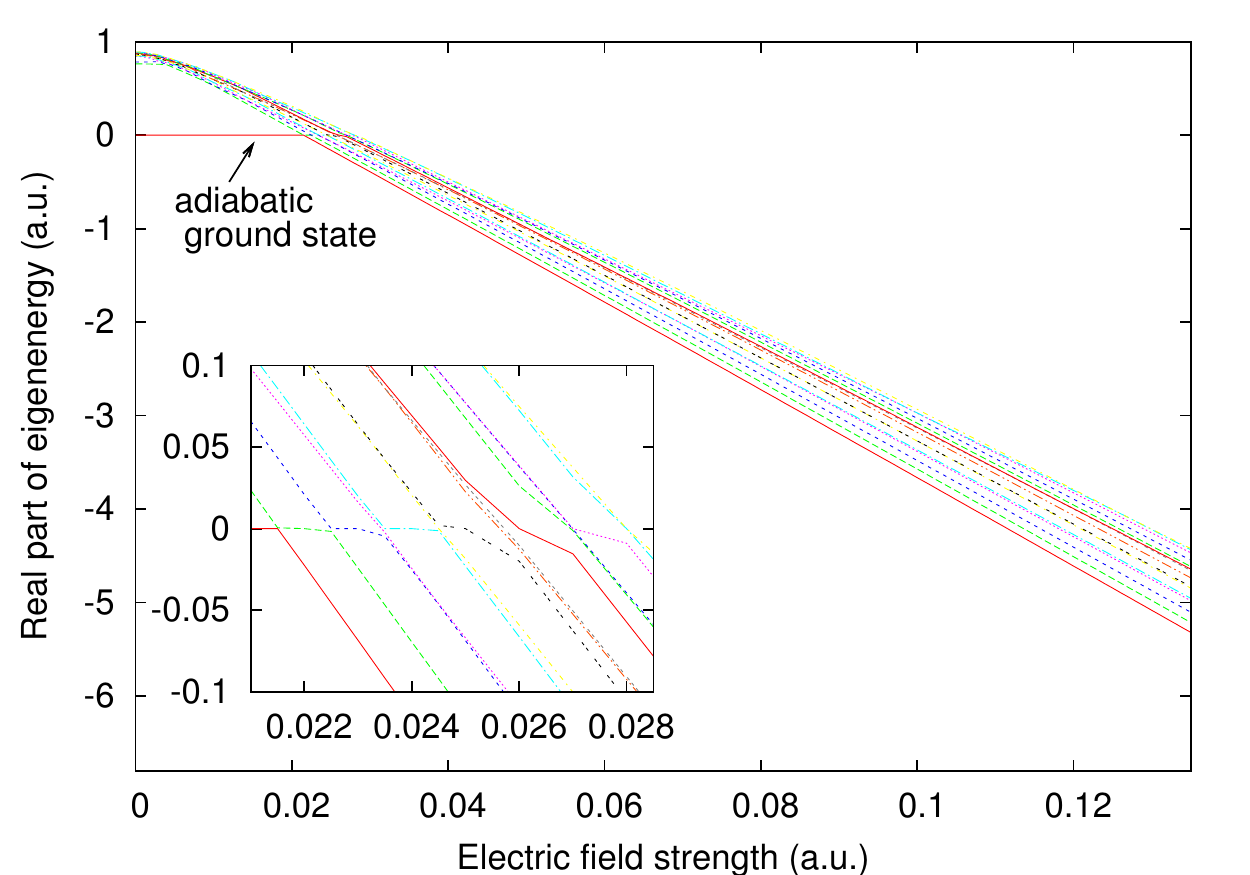}
\caption{The real part of the energy of several lowest lying adiabatic eigenstates as a function 
of the applied electric field strength~\cite{antonia1}. For small electric fields avoided crossings are observed (inset). Reprinted with permission from Ref.~\cite{antonia1}, \copyright ~2013 by the American Physical Society.}
\label{fig2}
\end{figure}

\begin{figure}[!tb]
\centering
\includegraphics[width=.75 \linewidth]{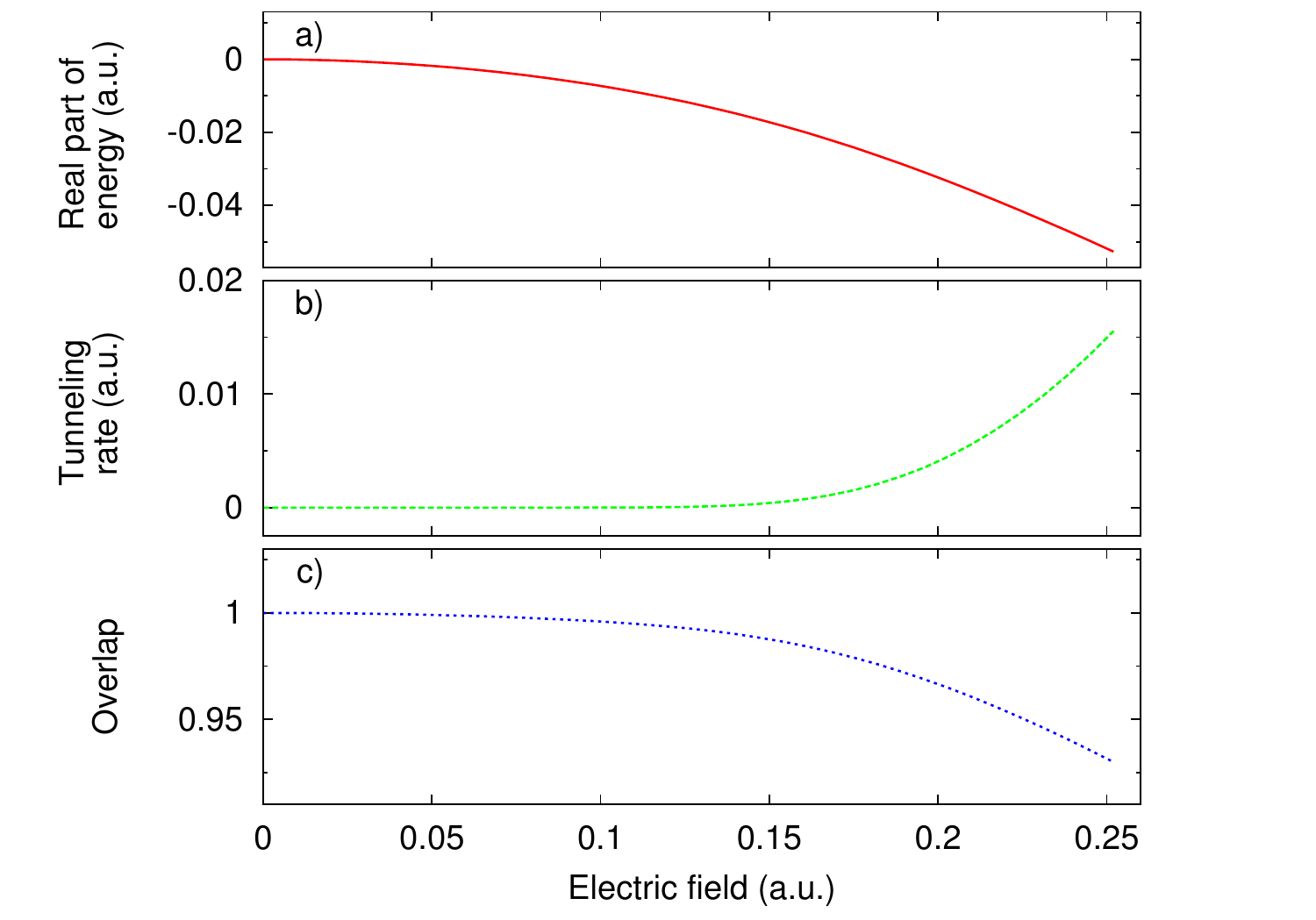}
\caption{(a) Real part of the energy of the constructed diabatic state, and (b) its tunneling rate shown as a function of the electric field~\cite{antonia1}. (c) Overlap of the diabatic state with the field-free ground state. Reprinted with permission from Ref.~\cite{antonia1}, \copyright ~ 2013 by the American Physical Society.}
\label{fig3}
\end{figure}

Solving Eq.~(\ref{adequation}) including a \acs{CAP} in the Hamiltonian for various field strengths yields the adiabatic eigenstates of the atom and the corresponding (complex) eigenenergies 
as a function of the field strength. The real part of the Siegert energies are shown in Fig.~\ref{fig2}. 
Many avoided crossings among the higher adiabatic eigenstates can be observed for field strengths 
in the range below $0.01$~a.u. ($1$~a.u.$= 5.14\times 10^{9}$~V/cm). The ground state energy does not change significantly up to field strengths around $0.02$~a.u.
 As shown in Ref.~\cite{antonia1}, the avoided crossings around $0.02$~a.u. (see inset of Fig.~\ref{fig2}) are simply jumped over and the system does not follow the adiabatic ground state (lowest red curve). Rather, for each applied field strength,  the electronic state follows the instantaneous eigenstate that has the maximal overlap with the field-free ground state. Therefore, in order to find a more meaningful description of the electronic state a diabatization method is employed whereby the diabatic state $|\Psi_0^{(d)}(t)\rangle$ is constructed from the adiabatic basis $\left\{ |\Psi_n(t)\rangle\right\}$ by requiring a maximal overlap with the field-free state $| \Psi^{(0)}\rangle$. To put it mathematically, for each field strength we seek the adiabatic state $|\Psi_n(t)\rangle$, for which $|\langle \Psi_n(t)| \Psi^{(0)}\rangle| >| \langle \Psi_m(t)| \Psi^{(0)}\rangle|$ for all $m \neq n$. Then, the diabatic state $ |\Psi_0^{(d)}(t)\rangle$ is set equal to $|\Psi_n(t)\rangle$. As can be seen in Fig.~\ref{fig3}\,c) the diabatic state has a $>90\%$ overlap with the field-free ground state. Figures~\ref{fig3}\,a) and \ref{fig3}\,b) show the real part of the energy and the tunneling rate as a function of the applied electric field. 
The quadratic behavior of the real part of the energy reflects the Stark effect and allows to read off the polarizability of the system. As expected, the tunneling rate increases considerably for sufficiently high field strengths, which can be understood within the picture visualized in Fig.~\ref{tunnelmulti}~a): The barrier width decreases because of the stronger field and thereby the tunneling probability is enhanced. 
The fact that the system follows one single diabatic state, which essentially maintains the character of the field-free ground state, gives a clearer and more intuitive picture for the explanation of the physics in the tunneling regime which, so far, was studied in the static case.
 
Furthermore, the diabatic state can be investigated for a dynamic scenario~\cite{antonia1} which involves a pulse with a Gaussian envelope:
\begin{equation}
 F(t) = f(t)\,\cos(\omega t) = F_0\,e^{-2\ln2t^2/\tau ^2}\,\cos(\omega t),
\end{equation}
where $F_0$ is the peak field strength, $\tau$ is the pulse duration (full width at half maximum, \acs{FWHM}) and $\omega$ is the field frequency. 

\begin{figure}[!t]
\centering
\includegraphics[width=.75\linewidth]{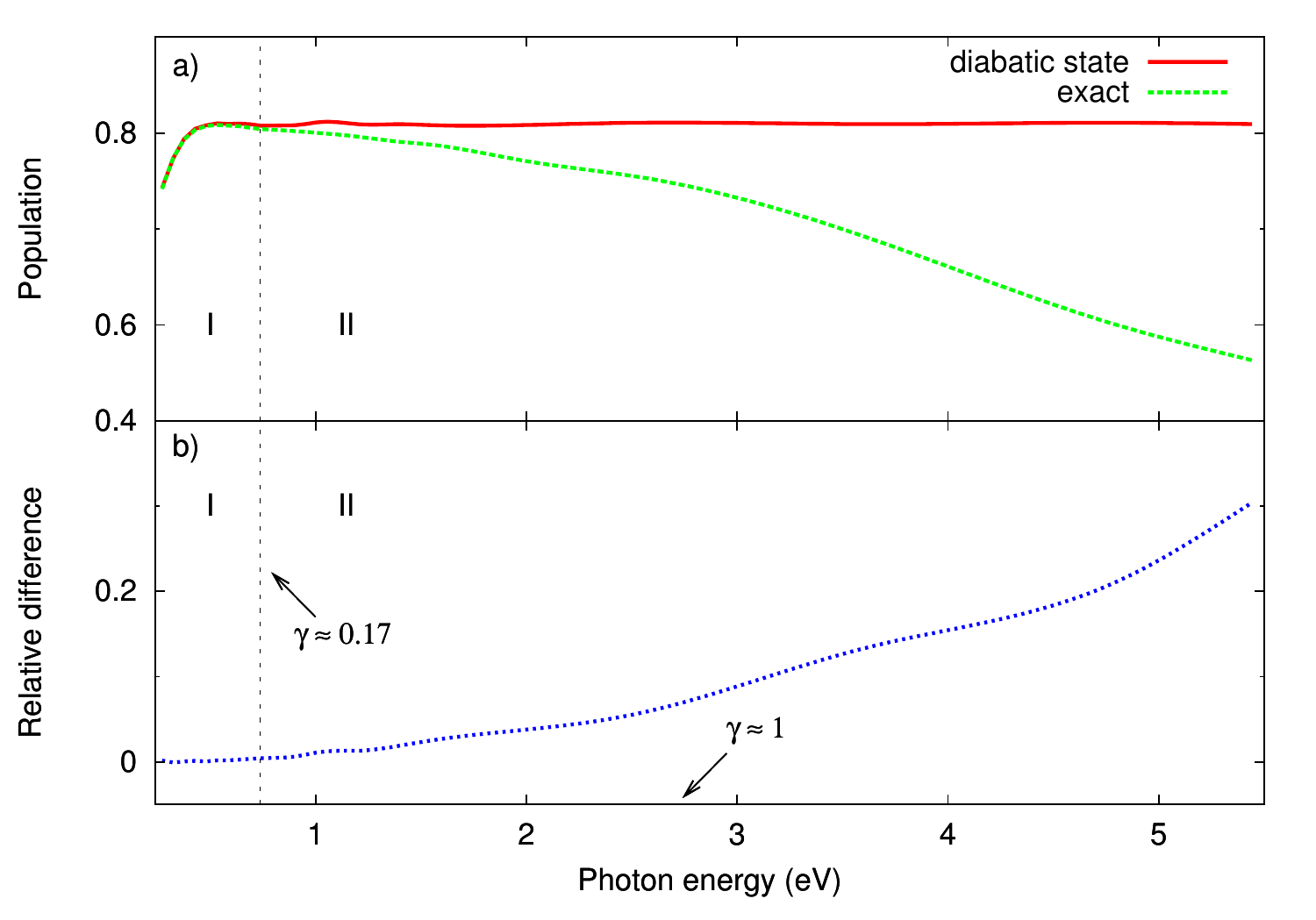}
\caption{a) Ground-state population after a Gaussian pulse with a peak field strength of $F_0=0.2$~a.u., and pulse duration of $400$~a.u.~\cite{antonia1}. The numerical solution of the Schr\"odinger equation is compared to the population obtained from the diabatic ground state as a function of the photon energy. b) Relative difference between the two results, hinting at the onset of nondiabaticity of the ionization process.  
The corresponding Keldysh parameter $\gamma$ is shown for different regions. Reprinted with permission from Ref.~\cite{antonia1}, \copyright ~2013 by the American Physical Society.}
\label{fig5}
\end{figure}

We wish to calculate the ionization probability for the diabatic state $|\Psi_0^{(d)}\rangle$ when applying this pulse. Moreover, we want to show that this state is sufficient to describe the dynamics in the tunneling regime because the system remains to a large extent in this state. Assuming that we have found the appropriate diabatic basis this particular diabatic state can be described in the diabatic basis by a single coefficient $\alpha_0^{(d)}$. Recalling Eq.~\eqref{eomadiab}, we can repeat the analysis for the diabatic case, and, neglecting transitions to other diabatic states (i.e., now making a "diabatic approximation"), we obtain the following equation of motion for the coefficients:
\begin{equation}
 i\dot{\alpha}_0^{(d)}(t)= \left[E_0^{(d)}-i\frac{\Gamma_0^{(d)}}{2}\right]\alpha_0^{(d)}(t),\label{diabeq}
\end{equation}
where $\Gamma_0^{(d)}$ is the decay rate of the diabatic state. The population evolution of our distinguished diabatic state, $P_0^{(d)}(t)=|\alpha_0^{(d)}(t)|^2$, during the pulse can be calculated with the equation of motion by taking the time derivative
\begin{equation}
 \frac{dP_0^{(d)}}{dt}=\frac{d}{dt}|\alpha_0^{(d)}(t)|^2=\alpha_0^{(d)*}(t)\dot{\alpha}_0^{(d)}(t)+\dot{\alpha}_0^{(d)*}(t)\alpha_0^{(d)}(t)= -\Gamma_0^{(d)}[F(t)] \  P_0^{(d)}(t),
 \end{equation}
where, in the last step, Eq.~\eqref{diabeq} was inserted. Of course, as we saw above, the ionization rate depends on the external field. 
This equation can easily be solved analytically by separation of variables:
\begin{equation}
 P_0^{(d)}(t) = \exp\left\{-\int_{-\infty}^{t} \dd t'\ \Gamma_0^{(d)}[F(t')]\right\}, \label{rate}
 \end{equation}
with the initial condition $P(t\!=\!- \infty)\! = 1$, i.e., all population being in this state long before the pulse. Since Eq.~\eqref{rate} accounts only for the population in our distinguished diabatic state, deviations in the population dynamics can be attributed to \textit{nondiabatic} behavior, i.e., transitions to other diabatic states. Therefore, we can compare the solution of this equation, in the diabatic approximation giving only the time evolution of our diabatic state, to the solution of the Schr\"odinger equation, which takes into account all different states and their transitions.
The result of this comparison for different photon energies is shown in Fig.~\ref{fig5} for a rather large peak field strength of $0.2$ a.u. Depicted are the populations calculated from Eq.~\eqref{rate} and the Schr\"odinger equation \eqref{tdse}, respectively, [Fig.~\ref{fig5}\,a)] and the relative difference [Fig.~\ref{fig5}\,b)] between them after the end of the pulse.
For sufficiently low energies, i.e., deep in the tunneling regime, the total ionization probability is reproduced exactly by considering only the 
diabatic state (region I). This means that the system remains in this single state and, therefore, it suffices for the description of the dynamics.
In the language of the adiabatic representation, the tunneling regime, where the Keldysh parameter $\gamma\ll 1$, is governed by a single diabatic state.
For higher photon energies around $1$~eV (region II), the difference increases significantly. In the region where $\gamma\approx 1$ the relative difference of the two methods amounts already to $\approx10\%$, which is a clear sign of transitions to other diabatic states. This indicates that nondiabatic effects start to become important. 

\begin{figure}[!tb]
\centering
\includegraphics[width=.75\linewidth]{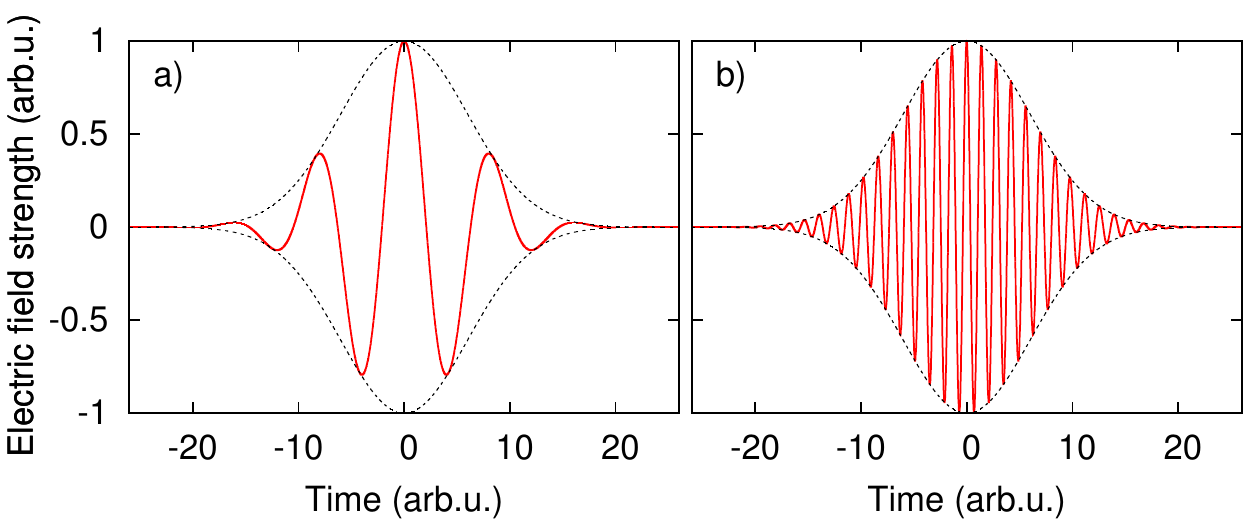}
\caption{Gaussian pulse $F(t)=\exp^{-2\ln2t^2/\t^2}\cos(\w t)$, the black-dotted line depicts the pulse envelope: a) few-cycle pulse, $\w=0.75$~arb.u., $\t=10$~arb.u., and b) multi-cycle pulse for the same pulse parameters, except $\w=4.5$.}
\label{fewmulti}
\end{figure}

Instead of varying the photon energy for a fixed pulse duration we can also divide the frequency range according to the number of cycles in the pulse. For the highest energies that are presented here multi-cycle pulses are considered [cf. Fig.~\ref{fewmulti}b)]. Going to lower photon energies, around $\approx0.8$~eV,we reach the few-cycle pulse regime [cf. Fig.~\ref{fewmulti}a)], where nice agreement of the diabatic state ionization and the solution of the Schr\"odinger equation is obtained. In contrast, tunneling theories such as the \acs{ADK} theory of tunneling ionization and similar approaches cannot reproduce the correct (diabatic) ionization rate for few-cycle pulses \cite{adk}.

In order to understand this difficulty we observe that the main difference between few- and multi-cycle fields is the change in field strength between consecutive field oscillations. For a few-cycle pulse the pulse envelope -- and with it the ionization rate --  changes significantly from one field oscillation to the next. This is depicted in Fig.~\ref{fewmulti}~a). As can be seen in Fig.~\ref{fewmulti}~b), the more cycles there are within the pulse, the less the variation in amplitude between consecutive oscillations.
In the framework of \acs{ADK} theory the ionization rate $\overline{\Gamma}(t)$ 
is obtained by integrating over one period of the field \cite{bis}
\begin{equation}
 \overline{\Gamma}(t) = \frac{1}{2\pi} \int_0^{2\pi} {\dd\varphi\ \Gamma[f(t)  \cos \varphi]}, 
\label{aver}
\end{equation}
where $\Gamma[F]$ is the instantaneous ionization rate.  
In the few-cycle limit the ionization rate cannot be simply averaged over one period, because the pulse envelope changes dramatically within one cycle. Of course, for multi-cycle pulses the rate can be averaged over the pulse and is then obtained from Eq.~(\ref{rate}) as
\begin{equation}
P_0^{(d)}(t) \approx \exp\left\{ -\int_{-\infty}^{t} \dd t'\ \overline{\Gamma}[{f}(t')]\right\}.
\end{equation}
Furthermore, we observe that the ionization probability in region I of Fig.~\ref{fig5} is not constant 
as a function of photon energy although it is well described by the ionization out of $|\Psi_0^{(d)}\rangle$ only. 
According to our analysis, this pronounced frequency dependence in the few-cycle limit should be attributed to the form of the pulse or, equivalently, to the relation between the number of cycles and the pulse envelope.

In this example we learn by employing the adiabatic representation that the Keldysh parameter is an approximate measure of diabaticity. In the few-cycle tunneling regime a single diabatic state is sufficient to describe the ionization dynamics. 

So far, we related the ionization dynamics to the eigenstates of the $N$-electron system. Let us now turn to the wave packet of the outgoing electron and the information it carries about the underlying photoionization process. To this end, a method for the calculation of photoelectron distributions is presented in the following.

\section{\label{sec:4} Calculation of photoelectron distributions within TDCIS}

Photoelectron spectroscopy has proven a powerful tool to analyze the processes that happen within complex systems upon irradiation and understand their electronic structure \cite{uwebecker}. Photoelectron spectra (\acs{PES}) and photoelectron angular distributions (\acs{PAD}) contain not only information about the interaction of the electrons with the light field, but also about the electronic correlations in the atomic shell~\cite{SchwarzElecSpectros80}. Photoelectron distributions can help to extract information and predict fundamental processes occurring during the interaction with the light pulse, e.g., to steer electrons with light waves \cite{kienberger}, which opened the way to attosecond streaking techniques \cite{drescher}, to study nonsequential and sequential double ionization in atoms \cite{nikolopoulos}, to scrutinize multielectron ionization dynamics \cite{bogus} and multiphoton excitations of deep shells of atoms \cite{meyerpapa}, and to reveal important information about the time-dependence of electron dynamics~\cite{kra,corkumkrausz}.

\subsubsection{Wave-function splitting method}\label{wavefunctionspl}
The calculation of the angular and energy-resolved photoelectron distribution allows for a direct comparison of theoretical predictions with experimental data. In principle, it can be done easily after the pulse is over by projecting the photoelectron wave packet onto the eigenstates of the field-free continuum, which corresponds to a Fourier transform of the wave packet. However, this approach requires large numerical grids in cases where the electrons can travel a long distance while the (possibly long and strong) pulse is still interacting with the system. Therefore, this approach can often not be applied to strong-field problems where the description of the ionized wave packet is particularly challenging due to the nonperturbative interaction between the electrons and the light pulse. 

For this reason new schemes, which overcome the obstacle of large grids, were developed for the calculation of photoelectron distributions. 
One method involves splitting the wave function into an internal and an asymptotic part \cite{tong} where the latter is then analyzed to yield the spectrum. Another method measures the electronic flux through a sphere at a fixed radius which allows to infer the parts of the outgoing wave packet \cite{scrinzi}. Both methods, which have been combined with \acs{TDCIS} \cite{PhysRevA.89.033415,PhysRevA.91.069907}, yield double-differential photoelectron spectra, i.e., the electron distribution as a function of kinetic energy and ejection angle. In the following the wave-function splitting method within \acs{TDCIS} is presented briefly~\cite{PhysRevA.89.033415}.

The wave-function splitting method rests on the assumption that once the ejected electron has traveled far enough from the ion it can be viewed as completely free and merely interacting with the laser pulse that might still be present. This is called the Volkov solution or strong-field approximation~\cite{PhysRevA.22.1786,Reiss2008}. Obviously, since in reality the Coulomb potential is a long-range potential the above-mentioned approximation will never be exactly true. Hence, further approximations treating the Coulomb field in higher order, like the Coulomb-Volkov approach were introduced~\cite{PhysRevA.18.538,refId0} and studies were performed in order to quantify the influence of the long-range potential on the physical observables~\cite{PhysRevA.75.043403}. However, often the potential's long-range character can be presumed to be negligible (e.g., fast electrons should be influenced only mildly at large distances), which renders this approximation valid.

To calculate the spectral components, the outgoing parts of the wave packet are split off the wave function of the system once they have reached a region far away from the atom. Since during the splitting procedure the laser field is still nonzero these parts must be analyzed using the Volkov Hamiltonian, which describes a free electron in the presence of an electric field~\cite{tong}. 

The theoretical framework of Sec.~\ref{theory} is employed to collect for each ionization channel all single excitations from the occupied spin orbital $|\varphi _i\rangle$ in one ``channel wave function'' by taking the sum over all virtual orbitals~\cite{roh}:
\begin{equation}
 |\chi_i(t)\rangle = \sum_a \alpha_i^a(t)|\varphi_a \rangle. \label{channelwfct}
\end{equation}
Using these channel wave functions all quantities are calculated in a channel-resolved manner. In this way, effectively, one-particle wave functions for each particular ionization channel $i$ are obtained.

\begin{figure}[!tb]
 \centering
 \includegraphics[width=.7\linewidth]{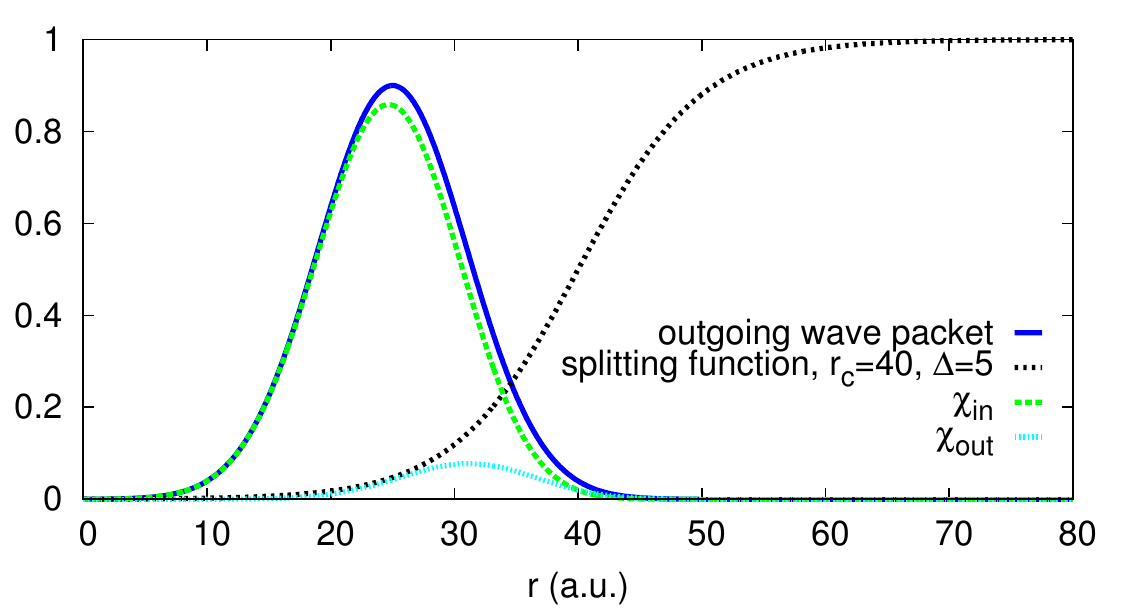}
 \caption{Schematic of the radial part of an outgoing wave packet (blue solid line), which is split by applying the splitting function (black dotted curve), as a function of distance to the ion. The inner (green dashed line) and the outer (light blue dash-dotted line) parts of the wave packet are also shown.}
 \label{splitwfct}
\end{figure}

As the method's name suggests, the channel wave function~\eqref{channelwfct} is split smoothly 
by applying the radial splitting function~\cite{tong}
\begin{equation}\label{splitfunction}
 \hat{S}= \left[1+e^{-(\hat{r}-r_c)/\D}\right]^{-1}.
\end{equation}
The parameter $r_c$ determines the center of the splitting function, and $\D$ is a ``smoothing'' parameter controlling the slope of the function. The splitting radius must be chosen sufficiently large, such that the electron is already far enough away to not return to the ion. The splitting function must not alter the ground state, i.e., $\hat{S}|\Phi_0\rangle=0$. This is achieved by requiring that $r_c/\Delta\gg 1$.
During the time propagation of the wave function the splitting is applied at certain splitting times $t_\mathrm{spl}$. At the first splitting time $t_0$ the channel wave function is split into two parts:
\begin{equation}
 |\chi_i(t_\mathrm{spl})\rangle = (1-\hat{S})|\chi_i(t_\mathrm{spl})\rangle + \hat{S} |\chi_i(t_\mathrm{spl})\rangle \equiv |\chi_{i, \mathrm{ in}}(t_\mathrm{spl})\rangle + |\chi_{i, \mathrm{ out}}(t_\mathrm{spl})\rangle.
\end{equation}
As visualized in Fig.~\ref{splitwfct}, when the splitting is applied $|\chi_{i,\mathrm{ in}}(t_\mathrm{spl})\rangle$ is the wave function in the inner region $0<r \lesssim  r_c$ and $|\chi_{i,\mathrm{out}}(t_\mathrm{spl})\rangle$ is the wave function in the outer region $r_c\lesssim r\le r_{\mathrm{max}}$.
We assume that far from the ion the electron experiences only the laser field and not the Coulomb field of the parent ion. Therefore, the outer part of the wave function $|\chi_{i,\mathrm{ out}}(t_\mathrm{spl})\rangle$ can be propagated analytically to a long time $T$ after the laser pulse is over using the Volkov Hamiltonian:
\begin{equation}
\hat{H}_V(\tau)=\frac{1}{2} \left[ \hat{\mathbf{p}}+\mathbf{A}(\t) \right]^2,
\end{equation} 
which involves the vector potential of the electric field. The corresponding Volkov time propagator has the form
\begin{equation}
 \hat{ U}_V(t_2,t_1)=\exp\left(-i \int_{t_1}^{t_2}  \hat{H}_V(\tau) \dd \t \right) \label{pes:volkovham}.
\end{equation}

The inner part of the wave function $|\chi_{i, \mathrm{ in}}(t_\mathrm{spl})\rangle$ still experiences the Coulomb couplings of the parent ion, and must be propagated using the full \acs{CIS} Hamiltonian [see Eqs.~\eqref{cis3} and \eqref{cis4}]. 

At each splitting time $t_\mathrm{spl}$ the outer part of the wave function is extracted. The inner part is propagated with the full Hamiltonian to the next splitting time and subsequently split into inner and outer part, and so forth, until all parts of the ejected electron wave packet have reached the outer region. Each outer part of the wave function is propagated analytically to large times after the pulse.

In order to calculate the spectral components of the emitted electron's wave packet, $|\chi_{i,\mathrm{ out}}(t_n)\rangle$ is expressed in the \acs{CIS} basis using new expansion coefficients $\beta_i^a(t_n)=\langle \varphi_a|\hat{S}|{\chi}_i(t_n)\rangle$ such that wave function in the outer region reads~\cite{PhysRevA.89.033415}
\begin{equation}
|\chi_{i,\mathrm{ out}}(t_n)\rangle=\sum_a\beta_i^a(t_n)|\varphi_a\rangle. \label{chiout}
\end{equation}

Employing the splitting method the Eqs.~\eqref{spatialeom} have the following form for the inner and outer part of the wave function, respectively~\cite{PhysRevA.91.069907}:
\begin{subequations}\label{pes:eom}
\begin{align}
 i \dot{\alpha}_i^a &= (\varepsilon_a-\varepsilon_i)\alpha_i^a +\! \sum_{i'b}\alpha_{i'}^b(2v_{ai'ib}-v_{ai'bi})+A(t)\!\left( \sqrt{2}\alpha_0\, p_{ai}+\!\sum_b p_{ab}\,\alpha_{i}^b 
-\!\sum_{i'}p_{i'i}\,\alpha_{i'}^a \right),\label{inner}\\
i \dot{\beta}_i^a &= (\varepsilon_a-\varepsilon_i)\beta_i^a +A(t)\left( \sum_b p_{ab}\,\beta_{i}^b-\sum_{i'}p_{i'i}\,\beta_{i'}^a\right).\label{outer}
\end{align}
\end{subequations}
The term involving $(2v_{ai'ib}-v_{ai'bi})$ in Eq.~\eqref{inner} vanishes for large distances as $1/r$~\cite{roh}, and, therefore, the Volkov approximation for the outer part of the wave function is justified. The last term in Eq.~\eqref{outer} couples different channel indices, $i,i'$, which introduces channel mixing in the outer part of the wave function.
In order to solve this system of differential equations we first observe that the time evolutions of the ionic and the electronic part of the electron wave packet in the outer region are decoupled and can be propagated with two time evolution operators to a large time $T$. Calling the ionic operator $\hat{U}^{\mathrm{ion}}(T,t)$ and the electronic Volkov time propagator $\hat{U}^{\mathrm{elec}}(T,t)=\hat{U}_V(T,t)$, see Eq.~\eqref{pes:volkovham}, we can write the time evolution for the coefficients of the outer part of the wave function as~\cite{PhysRevA.91.069907}
\begin{equation}
\beta_i^a(T;t_n) =\sum_b \hat{U}^{ab}_V(T,t_n)\sum_j \hat{U}_{ij}^{\mathrm{ion}}(T,t_n) \beta_j^b(t_n)=\sum_b \hat{U}_V^{ab}(T,t_n)\beta_i^b(t_n).\label{betacoeff}
\end{equation}
Inserting this into Eq.~\eqref{outer} yields the following equation of motion for the ionic time evolution operator:
\begin{subequations}\label{eomuion}
\begin{align}
i\pa_t U_{ij}^{\mathrm {ion}}(T,t)&=\left[-  \varepsilon_i \delta_{ij}U_{ij}^{\mathrm {ion}}(T,t) - A(t)\sum_{i'} \langle \varphi_{i'}| \hat{p}| \varphi_i\rangle U_{i'j}^{\mathrm {ion}}(T,t) \right] ,
\end{align}
\end{subequations}
with $U_{ij}^{\mathrm {ion}}(t,t)=\delta_{ij}$. It is solved by numerical time propagation (e.g., using the Lanczos-Arnoldi algorithm) up to large times $T$ and thereby the coefficients $\beta_i^b(t_n)$ are obtained.

Now, we need to calculate the spectral components. The Volkov states $ |\mathbf{p}\,^V\rangle=(2\p)^{-3/2}e^{i\mathbf{p}\cdot\mathbf{r}}$, which in the velocity form are plane waves, are the eigenstates of the Volkov Hamiltonian and form a basis set in which the channel wave packet for each channel $i$ at final time $T$ can be expanded:
\begin{equation}
 |\chi_{i,\mathrm{ out}}(T)\rangle= \int\! \dd^3p \sum_{t_n}C_{i}(\mathbf{p},t_n) \,| \mathbf{p}\,^V\rangle,\label{pes:coeff}
\end{equation}
where a sum of the spectral components over all splitting times $t_n$ must be performed. Inserting Eq.~\eqref{chiout}, the corresponding coefficients are obtained at final time $T$ as
\begin{equation}
 C_i(\mathbf{p},t_n)= \langle  \mathbf{p}\,^V|\hat{ U}_V(T,t_n)\left(\sum_a\beta_i^a(t_n)|\varphi_a\rangle\right) .
\end{equation}

Then, an incoherent summation over all possible ionization channels must be performed to obtain the double-differential photoelectron distribution as a function of the kinetic energy and the angle with respect to the light polarization axis (because of the linear polarization the distribution is symmetric in the azimuthal angle):
\begin{equation}
 \fr{d^2P(\mathbf{p})}{dE d\W}=p \sum_i \bigg| \sum_{t_n} C_i(\mathbf{p},t_n\, ) \bigg|^2.\label{pes:pespectrum}
\end{equation}
The extra factor of $p$ results from the conversion from the momentum to the energy differential. The final time $T$ must be chosen sufficiently large, in order to ensure that all parts, in particular the low-energy components, of the outgoing wave packet have reached the outer region and can be analyzed. Further details can be found in Ref.~\cite{PhysRevA.89.033415}.


\section{\label{sec:5} Applications}
In the following a few examples of multiphoton ionization of atoms will be presented. In particular we choose the process of above-threshold ionization (\acs{ATI}) for the investigation in different photon energy regimes. \acs{ATI}, first observed in 1979 by P. Agostini {\normalfont \itshape et al.} \cite{PhysRevLett.42.1127}, is a highly nonlinear process where an electron absorbs more photons than are necessary for ionization~\cite{gontier}. As a consequence, a series of peaks can be observed in the photoelectron spectrum, where the separation between two consecutive peaks corresponds to the energy of one photon.

\subsection{\acs{ATI} in the \acs{XUV} energy range}
Let us employ the wave-function splitting method to understand the \acs{ATI} of argon ($Z=18$, closed shell $1s^22s^22p^63s^23p^6$) in the \acs{XUV} energy range. 
Numerically, for the splitting method to work properly, three parameters have to adjusted: the splitting radius, $r_c$, the smoothness of the splitting function, $\D$ [cf. Eq.~\eqref{splitfunction}], and the time instances at which the absorption is applied during the time propagation, $t_{\rm spl}$. 
In the following we apply a strong Gaussian \acs{XUV} pulse centered at $\w=105$~eV with $\D\w=0.7$~eV bandwidth (\acs{FWHM}), which corresponds to a Fourier-transform limited pulse with $\t=108$~a.u. ($2.6$ fs) duration (\acs{FWHM}).\footnote{A pulse having the minimum possible pulse duration given an energy bandwidth is called Fourier-transform limited and the relation holds (in atomic units): $\Delta\w\D\t=2.765$.} The peak intensity of the pulse is $1.0\times 10^{15}$~Wcm$^{-2}$. At this photon energy the $3s$ and the $3p$ electrons in argon can be ionized with one photon. If the pulse intensity is high enough they can absorb two photons and undergo \acs{ATI}.

\begin{figure}[!tb]
 \centering
 \includegraphics[width=\linewidth]{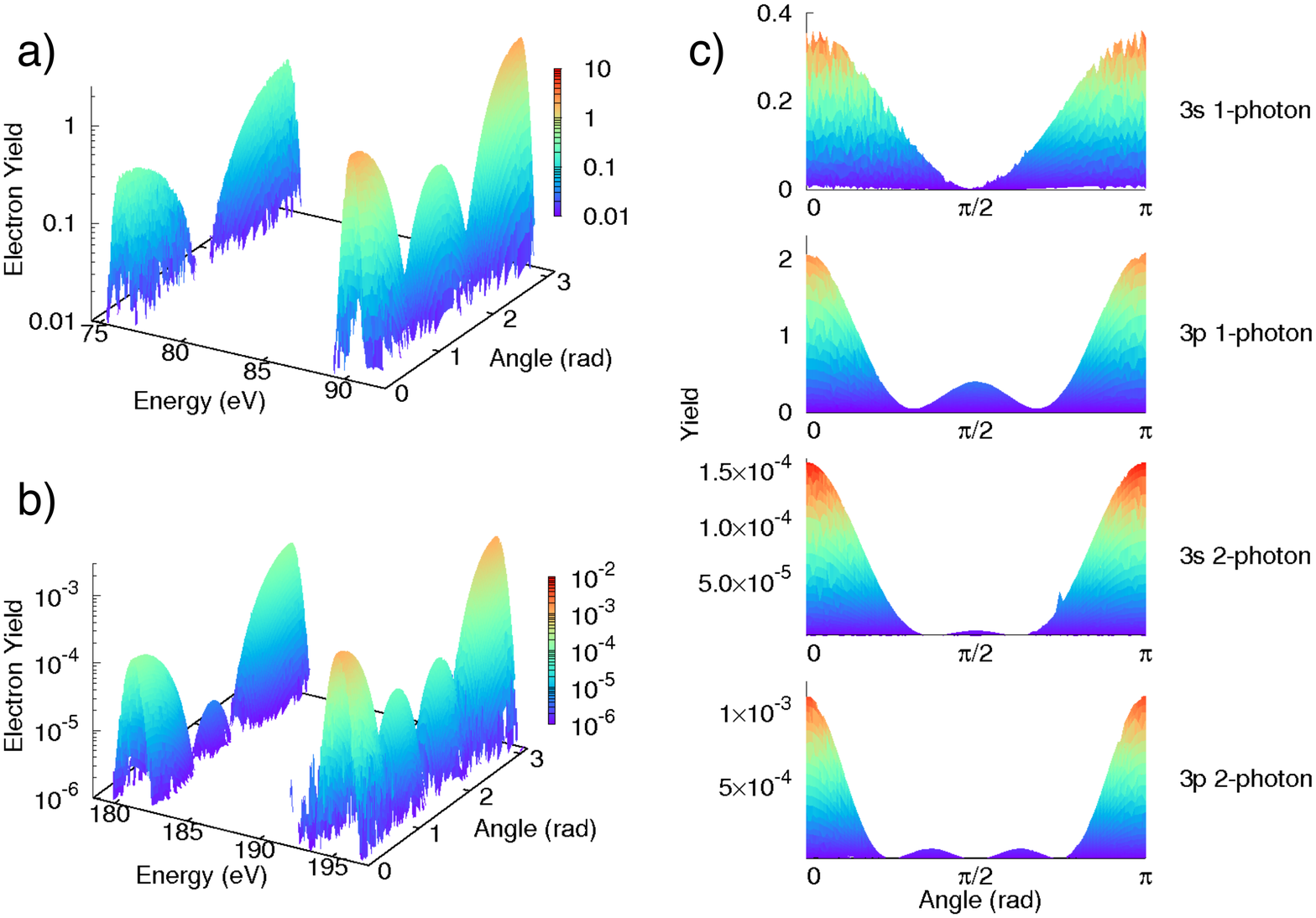}
 \caption{Energy- and angle-resolved argon photoelectron distribution produced with the splitting method is shown for an \acs{XUV} pulse at $105$~eV photon energy, $10^{15}$~Wcm$^{-2}$ intensity and $2.6$ fs pulse duration~\cite{PhysRevA.89.033415}. The angle denotes the direction with respect to the polarization axis of the pulse. The angular distribution reflects the change in angular momentum by a) one- and b) two-photon absorption. c) Energy cuts of the angular distribution profile at the peak maxima. Reprinted with permission from Ref.~\cite{PhysRevA.89.033415}, \copyright~2014 by the American Physical Society.}
 \label{fig4}
\end{figure}

First we present the double-differential photoelectron distribution, i.e., the ionization yield as a function of energy and angle~\cite{PhysRevA.89.033415}. In Fig.~\ref{fig4}~a) the full angle- and energy-resolved photoelectron distribution of argon after one-photon absorption is shown.
The energies of the peaks correspond to the difference between the photon energy and the binding energy of the corresponding orbital ($3s$ and $3p$, respectively). The energy width of the peaks corresponds to the Fourier-transform limited energy width. 
 Fig.~\ref{fig4}~b) shows the corresponding \acs{ATI} peaks, which are separated in energy from the one-photon peaks exactly by the energy of one photon. 
 
The second axis represents the angle with respect to the polarization axis. The angular distributions, which are different for the different peaks, feature the corresponding contributions from the different channels. They are visualized in Fig.~\ref{fig4}~c) as cuts along the maxima of the corresponding peaks. In one-photon ionization the $3s$ electron occupies a $p$ state ($l=1$) which manifests itself in the $Y_{10}\propto\cos^2\theta$ distribution. The $3p$ one-photon peak has both an $s$- and a $d$-wave contribution. Analogously, through the absorption of two photons the \acs{ATI} peak of the $3s$ electron exhibits an $s$- and $d$-wave character and the $3p$ peak a $p$- and $f$-wave character.

\begin{figure}[!tb]
 \centering
 \includegraphics[width=.75\linewidth]{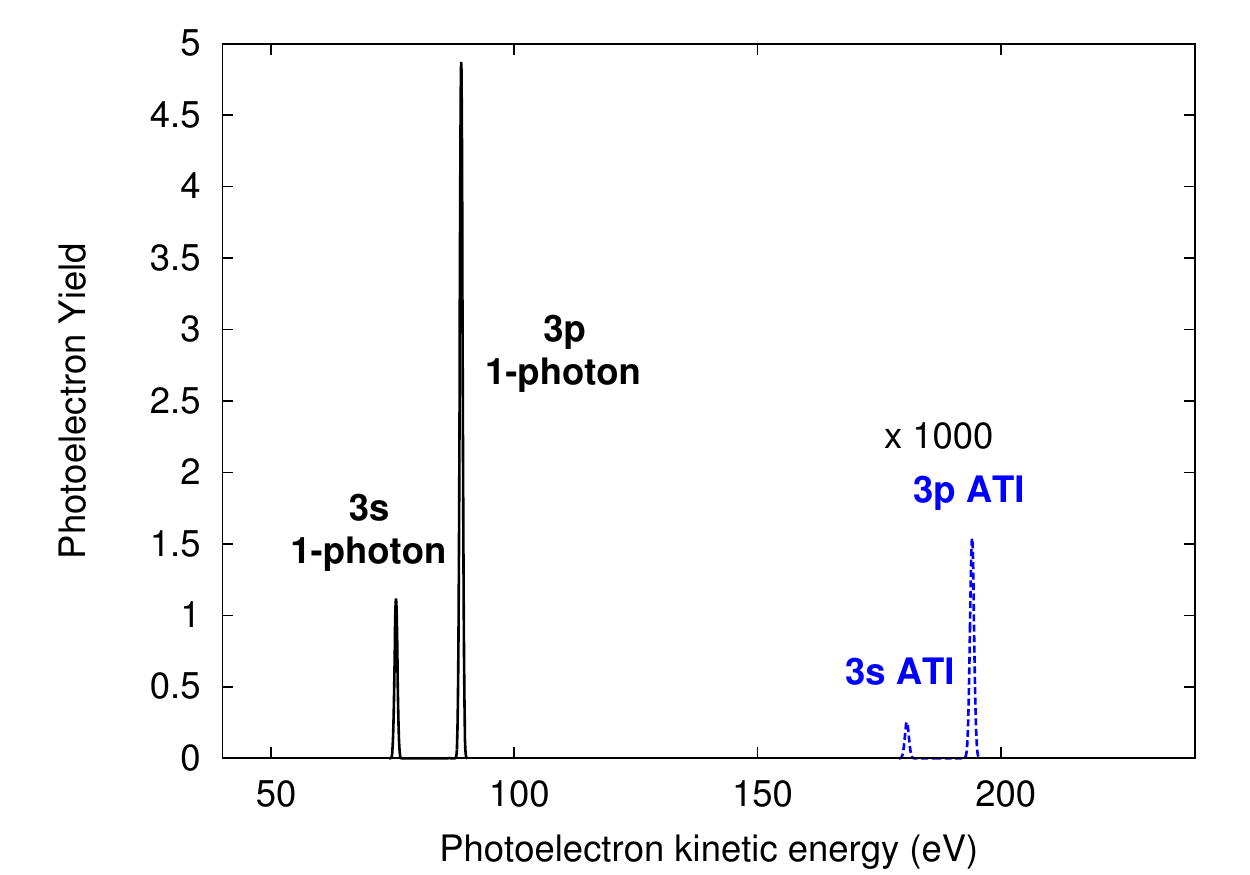}
\caption{Angle-integrated photoelectron distribution of argon, cf. Fig.~\ref{fig4}. The dominating part of ionized population stems from the $3p$ and the $3s$ shells, which absorb one photon (solid black lines). The \acs{ATI} peaks are shown as well, magnified by a factor of $1000$ for better visibility (dashed blue lines).} 
\label{arenergypes}
\end{figure}

When integrated over the solid angle the photoelectron distribution is just a function of the photoelectron's kinetic energy:
\begin{equation}
\fr{d P(\mathbf{p})}{dE}=\int_0^{2\p} d\phi\int_0^\p d \theta \sin\theta \fr{d^2P(\mathbf{p})}{dE d\W}.
\end{equation}
In that case it is commonly called photoelectron spectrum (\acs{PES}). In Fig.~\ref{arenergypes} the photoelectron spectrum corresponding to the photoelectron distribution of Fig.~\ref{fig4} is shown, i.e., integrated over the solid angle. The peaks reflect the binding energy position and allow for the characterization of the corresponding shell that was ionized. Such energy-dependent \acs{PES} are recorded in experiment when all electrons irrespective of their emission direction are measured, e.g., by a magnetic-bottle spectrometer. In the next subsection, this type of spectra measured for atomic xenon will be of central importance.

\subsection{Xenon \acs{ATI} involving the giant dipole resonance}
Now we will investigate a system exhibiting electron correlations in the atomic shell by employing multiphoton ionization as a tool.

In 1964 a strong response of xenon ($Z=54$) to \acs{XUV} radiation was discovered in the one-photon absorption spectrum \cite{ederer64}, the so-called giant dipole resonance (\acs{GDR}).
The three outermost shells of xenon are the $5p^6$, $5s^2$, and $4d^{10}$ shells. The \acs{GDR} was soon attributed to the interplay of two effects concerning the $4d$ shell. Qualitatively the phenomenon can be explained as originating from a shape resonance effect: After absorbing one photon a $4d$ electron is promoted to the continuum. Due to the centrifugal barrier (predominantly angular momentum $l=3$), it is trapped temporarily in a resonance state near the ionic core \cite{Coop64,Sta82} until it tunnels out and leaves the ion. The \acs{XUV} resonance is interpreted as the collective response of all ten $4d$ electrons~\cite{Fet71}. Only when electron correlation effects within the $4d$ shell are included is quantitative agreement with experimental data achieved. This means that the resonant excitation cannot be explained as a purely independent-particle effect. Until recently it was assumed that the \acs{GDR} consists of one single, broad resonance which is associated with one quantum state.
In this section it will be shown that using \acs{XUV} two-photon spectroscopy, there is substantial sensitivity to substructure of this resonance. In fact, the \acs{GDR} accommodates two resonances.
\begin{figure}[!tb]
\begin{centering}
 \includegraphics[width=.75\linewidth]{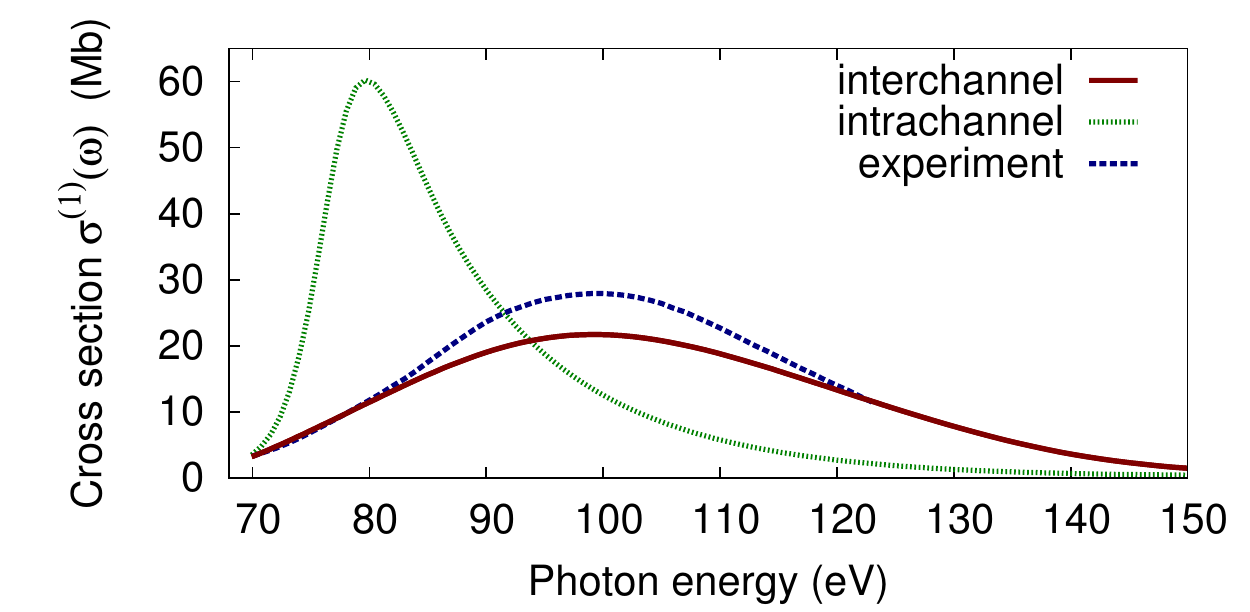}
 \caption{One-photon absorption cross section of xenon calculated within \acs{TDCIS} using the two models. The experimental curve \cite{Samson2002265} resembles the interchannel curve \cite{krebs14}.}
 \label{xeati:1phcs}
 \end{centering}
\end{figure}

The \acs{GDR} has been studied previously within \acs{TDCIS} \cite{Pab13,krebs14}. 
In Fig.~\ref{xeati:1phcs} 
the one-photon cross section of xenon is shown for the intrachannel and interchannel models. 
The interchannel model captures the main features of the many-body effect that renders the 
one-photon cross section curve broader and shifts it to higher energies. In fact, the interchannel curve reproduces reasonably well both the position and the width of the experimental cross section~\cite{krebs14}. 

\subsubsection{Nonlinear response regime}

\begin{figure}[!b]
 \includegraphics[width=\linewidth]{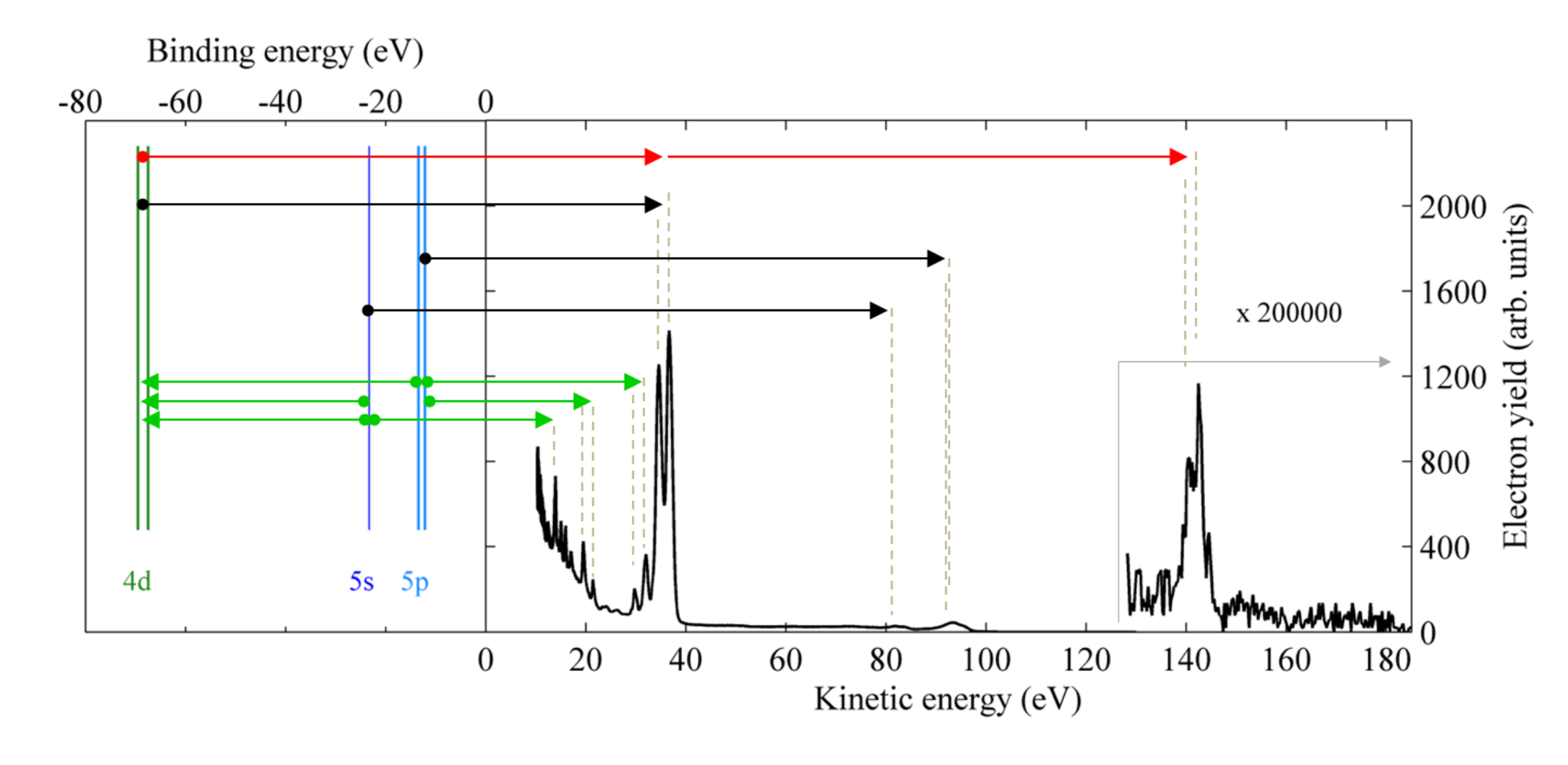}
 \caption{Electronic level scheme and photoelectron spectrum from \acs{XUV} ionized xenon atoms, recorded at $\hbar\w = 105$ eV at an intensity of $6 \cdot 10^{12}$ Wcm$^{-2}$~\cite{Maz15}. The spectrum includes features caused by different processes represented by arrows: one-photon direct emission (black), Auger emission (green) and two-photon emission (red). Reprinted with permission from Ref.~\cite{Maz15}, \copyright ~Nature Publishing Group}
 \label{xeati:fig2}
\end{figure}

\begin{figure}[!tb]
\begin{centering}
 \includegraphics[width=.75\linewidth]{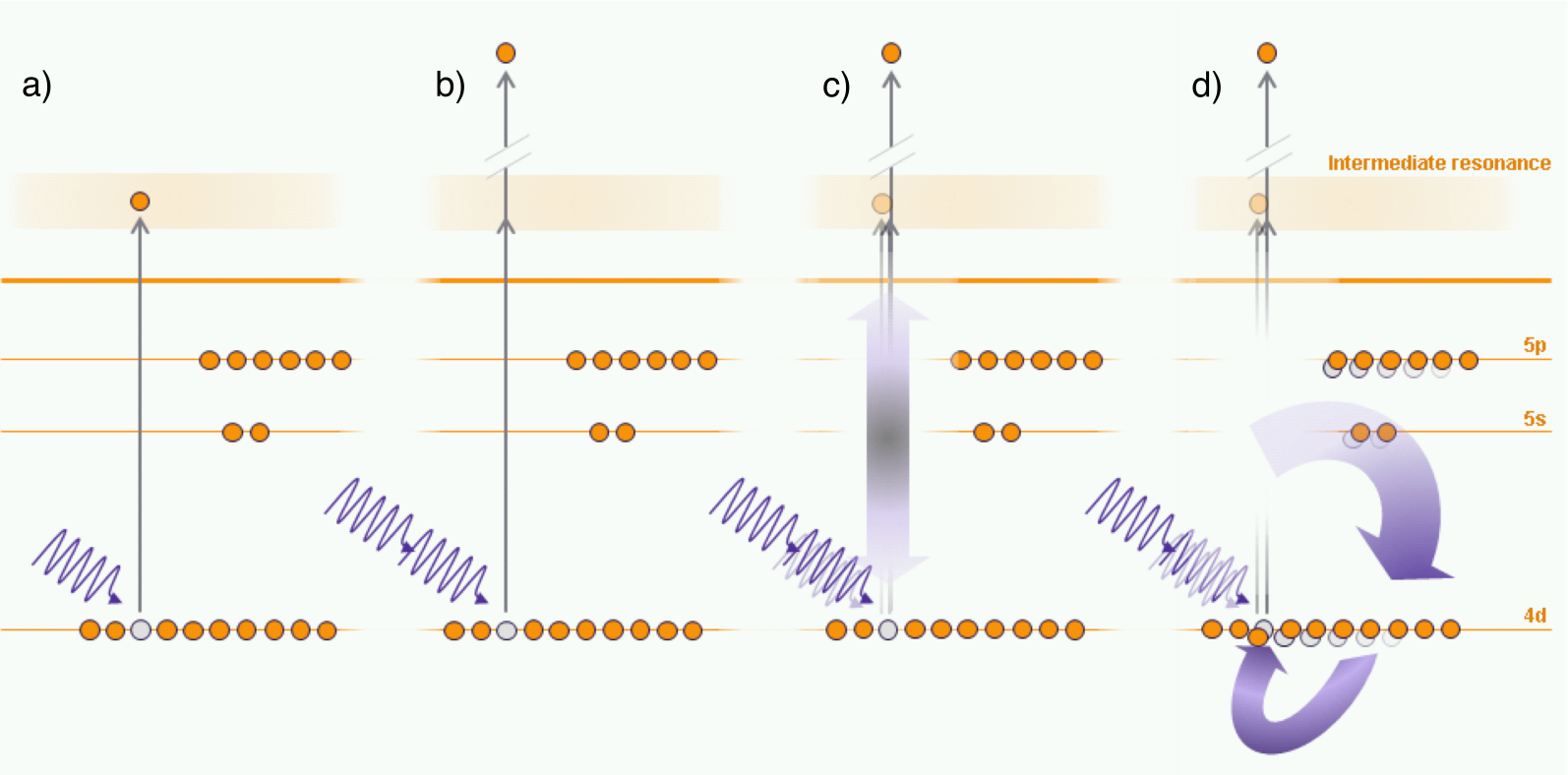}
 \caption{Schematic representation of the ionization processes~\cite{Maz15}. a) One-photon ionization; b) \acs{ATI}; c) one-photon ionization and \acs{ATI} in the intrachannel case; d) one-photon ionization and \acs{ATI} in the interchannel case, accounting for electron-hole interaction in all channels open to ionization. Reprinted with permission from Ref.~\cite{Maz15}, \copyright ~Nature Publishing Group}
 \label{xeati:fig1}
 \end{centering}
\end{figure}

Let us now review the analysis of the \acs{GDR} in the two-photon regime and the comparison of the theoretical results with experimental data obtained at the free-electron laser \acs{FLASH}, which was originally published in Ref.~\cite{Maz15}. In the experiment xenon gas was irradiated by an \acs{FEL} beam and the resulting photoelectrons were recorded with a magnetic-bottle electron spectrometer. As we saw above, photoelectron spectra allow to disentangle one-photon ionization from \acs{ATI}. One of the spectra that were recorded is shown in Fig.~\ref{xeati:fig2}. 
The processes of interest are the one- and two-photon ionization of the $4d$ shell. Since the photon-energies selected in the experiment lie exactly in the range of the \acs{GDR}, $105$~eV and $140$~eV, the two-photon process occurs through the giant dipole resonance as an intermediate step (Fig.~\ref{xeati:fig1}). Subsequently the inner-shell vacancy decays via Auger decay \cite{Sie72}. In this case the two-photon process is \acs{ATI} because the photon energies exceed the binding energy of the $4d$ orbital. In Fig.~\ref{xeati:fig1} the levels of xenon and the ionization processes are shown schematically. All of these processes can be identified in the experimental spectra, cf. Fig.~\ref{xeati:fig2}. At kinetic energies around $140$~eV, the two-photon ionization from the $4d$ shell is observed in a spectral feature which resembles the shape of the $4d$ (one-photon) emission lines around $35$~eV and is separated from them by exactly the energy of one photon.

In order to quantify the ionization processes the two-photon cross section is calculated within \acs{TDCIS} by calculating the depopulations in the $4d$ shell due to one- and two-photon absorption~\cite{pab}. As long as perturbation theory is valid and higher order processes (e.g., three-photon processes) are negligible, the depopulations are distinguishable due to the different angular momenta of the ejected electron according to the dipole selection rules. All contributions from the $4d_m$ subshells, with $m$ being the magnetic quantum number, i.e., $4d_0$, $4d_{\pm1}$, and $4d_{\pm 2}$, must be added. Equivalently, the corresponding peak of the photoelectron distribution can be integrated over the angle and the energy in order to obtain the ionized population.

In the following, the role of collectivity shall be investigated in the nonlinear regime. As described in Sec.~\ref{theory} a comparison between the interchannel and intrachannel models can elucidate the impact of collectivity. For the description of a collective response the system cannot be written as a single particle-hole state, but instead a superposition of particle-hole states is needed. The interchannel model includes the coupling among the holes in the $4d$, $5s$ and $5p$ orbitals and the electron, see Fig.~\ref{xeati:fig1}\,d), cf. also Fig.~\ref{intrainter}\,b). The corresponding Coulomb matrix elements $\langle \Phi_i^a|\hat{V}_{\mathrm {e-e}}|\Phi_j^b\rangle$ between the particle--hole excitations $|\Phi_i^a\rangle,\,|\Phi_j^b\rangle$, are included for all different hole pairs $(i,j)$ within the space of active orbitals ($4d$, $5s$, and $5p$), which means that $\langle \Phi_i^a|\hat{V}_{\mathrm {e-e}}|\Phi_j^b\rangle\neq 0$, for all $i=j$ {\em{and}} $i\neq j$; $a,b$ are taking values for all virtual orbitals. In this way, superpositions of particle-hole states, i.e. collective states, may be described. 
In contrast, in the case of the intrachannel model the elements with $i\neq j$ are set to zero. Keeping only the elements $\langle \Phi_{4d_{ m}}^a|\hat{V}_{\mathrm {e-e}}|\Phi_{4d_{ m}}^b\rangle\neq 0$ results, therefore, in the description of coupling only with the $4d_m$ orbital from which the electron was ionized, see Figs.~\ref{xeati:fig1}\,c), cf. also \ref{intrainter}\,a). The relevant cross sections are calculated for both schemes and in the following the results will be compared to the experimental data.

As introduced in Sec.~\ref{theory} rate equations can describe the ionization probability in the perturbative regime. The corresponding differential equations describing the ionization out of the $4d$ shell by a light pulse have the form:
\begin{subequations}\label{rateeq}
\begin{eqnarray}
 \fr{d P_0}{d t } & =& - \left[ p_{1ph}(t) + p_{2ph}(t) \right] ,\\
p_{1ph}(t) &=& \sigma^{(1)} \  j(t) \ P_0(t), \label{eq:rate}\\
p_{2ph}(t) &=& \sigma^{(2)} \ j^2(t) \ P_0(t) ,
\end{eqnarray}
\end{subequations}
where $P_0$ denotes the $4d$ ground state population. The first equation describes the change in population of the ground state due to one- and two-photon ionization, denoted by $p_{1ph}$ and $p_{2ph}$, respectively. $\sigma^{(1)} $ and $\sigma^{(2)}$ are the one-photon and \acs{ATI} cross sections, respectively. $j (t)=I(t)/\w$ is the time-dependent flux, where $I(t)$ is the pulse intensity and $\w$ is the photon energy. In other words, the flux is a measure of the number of photons that are available for ionization processes per unit time and area. 

The rate equation system \eqref{rateeq} is solved for a variety of pulse intensities spanning many orders of magnitude (e.g., using the Runge-Kutta algorithm). One- and two-photon ionization yields are obtained for both the intrachannel and interchannel model by inserting the corresponding cross sections. 

In our considerations the ground state population can be set constant, i.e., $P_0(t)=P_0=1$ because a crucial criterion for the perturbative regime is that the ground state is not depleted. Thereby the equations are simplified and can be integrated to yield the depopulations due to $N$-photon absorption
\begin{eqnarray}\label{popsandcs}
P_{N}= \int p_{Nph} dt=\sigma^{(N)} \int j^N (t)d t  \equiv \sigma^{(N)} F^{(N)},
\end{eqnarray}
Here, $F^{(N)}=\int j^N (t)d t$ denotes the fluence that is available for the ionization by $N$ photons. For all our purposes we will use Gaussian pulses with the intensity envelope
\begin{equation}
I(t)=\fr{cE_0^2}{8\p}\exp\left(-4\ln 2\  t^2/\t^2\right), \label{multi:gaussian}
\end{equation}
where $c$ is the speed of light, $E_0$ is the peak electric field, and $\t$ is the pulse duration (\acs{FWHM}). We obtain the fluences for one-photon and two-photon absorption:
\begin{subequations}\label{fluences}
\begin{eqnarray}
F^{(1)}&=&\fr{c\t}{8\p\hbar\w}\sqrt{\fr{\p}{4\ln 2}}E_0^2,\\
F^{(2)}&=&\left(\fr{c}{8\p}\right)\sqrt{\fr{\p}{8\ln 2}}\fr{E_0^4}{(\hbar\w)^2}.
\end{eqnarray}
\end{subequations}
From Eq.~\eqref{popsandcs} the expressions for the cross sections follow: $\sigma^{(N)}= P_{N}/F^{(N)}$. The units of (generalized) cross sections are cm$^{2N}$s$^{N-1}$ \cite{saenz}.

\subsubsection{Comparison between theory and experiment}

In order to meaningfully compare the theoretical calculations with experimental data it is necessary to perform a volume integration matching the experimental setting. Because of the experimental spatial pulse profile the light intensity varies over the interaction region of the light with the gas. Therefore, also the ionization probability depends on the position of an atom within the electric field distribution. To find the correct spatial dependence the following parameters are important: the duration and the statistics of the light pulses, the beam focus, the Rayleigh length, the volume of acceptance along the propagation direction and perpendicular to it, described by the coordinate $z$ and the radial coordinate $\rho$, respectively. 

To this end, we use our signal, $S(F)$, which is given as a function of intensity or, equivalently, of fluence. 
A Gaussian beam profile is assumed, i.e., the fluence distribution is given by $F(\rho,z)= F_{0}(z)\exp[-\rho^2/w(z)^2]$, with $F_{ 0}(z)=4n_{phot}\ln 2 /[\p w^2(z)]$. $n_{phot}$ is the number of photons in the pulse. The function $w(z)$ describes the divergence of the beam, i.e. it determines the spot size at a distance $z$ from the beam waist or focus $w_0$:
\begin{equation}
w(z)^2=w_0^2\left[ 1+\left(\frac{z}{z_0}\right)^2 \right].
\end{equation}
 Denoting the integration volume element by $\dd V$ we obtain the expression
\begin{equation}
 \int S(F)\dd V = \int \rho\ \dd \rho\ \dd z  S(F) = \int \dd z \ \dd F \ \rho(F,z)\big| J(F,z)\big|  S(F) ,\label{multi:signal}
\end{equation}
where a coordinate transformation $(\rho,z)\rightarrow(F,z)$ is performed with the determinant of the Jacobian transformation matrix
\begin{equation}
  J(F,z)= \begin{pmatrix} 
 \frac{\partial \rho}{\partial F}& \frac{\partial \rho}{\partial z} \\
  \frac{\partial z}{\partial F} & \frac{\partial z}{\partial z}
\end{pmatrix},
\end{equation}
because an integration over the fluence is more favorable than over the radial coordinate.
The fluence distribution can be inverted bijectively and yields $\rho(F,z)$ when we take the positive (physical) solution. We finally obtain
\begin{equation}
 \int S(F)\dd V=\int_0^{F_{0}} \dd F\ S(F) \left\{ \int_{z_\mathrm{min}}^{z_{\mathrm{max}}} \rho(F,z)|J(F,z)|\dd z \right\},
\end{equation}
which can be calculated numerically employing the experimental parameters.

\begin{figure}[!tb]
 \includegraphics[width=.75\linewidth]{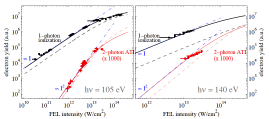}
 \caption{Intensity dependence of one-photon and \acs{ATI} yields of xenon~\cite{Maz15}. Experimental electron yields (full dots) as a function of \acs{FEL} intensity are shown at 105 eV and 140 eV photon energy. The dash-dotted lines show the linear and quadratic dependence of the one-photon and \acs{ATI} yields, respectively [cf. Eq.~\eqref{logdep}]. At higher intensities, saturation effects appear due to the depletion of the neutral atom. The lines represent theoretical yields in the interchannel (solid) and intrachannel (dashed) scheme, respectively. Reprinted with permission from Ref.~\cite{Maz15}, \copyright ~Nature Publishing Group}
 \label{xeati:fig3}
\end{figure}

In Fig.~\ref{xeati:fig3} the comparison between the two theoretical models, intrachannel and interchannel, and the experimental data is drawn for the two photon energies of $105$ and $140$~eV~\cite{Maz15}. It clearly shows that the interchannel model (solid lines) reproduces the intensity dependence of the experimental yields, whereas the intrachannel model (dashed lines) fails to do so. Note that the slopes of the dash-dotted lines represent the intensity dependence of a linear and a quadratic process according to Eq.~\eqref{logdep} and match the experimental points of the corresponding ionization order. The excellent agreement in the ratio between the two orders of ionization and the onset of saturation confirms that the interchannel model captures the relevant physics over the whole intensity range.
This leads to the conclusion that correlations between all possible electron-hole states play a major role in the two-photon ionization process.
 At $140$ eV the experimental results are described particularly well by the interchannel model which gives a much larger cross section than the intrachannel model.

\subsubsection{Theoretical analysis of the \acs{ATI} cross section}\label{theorycross}

\begin{figure}[!tb]
\begin{centering}
 \includegraphics[width=.75\linewidth]{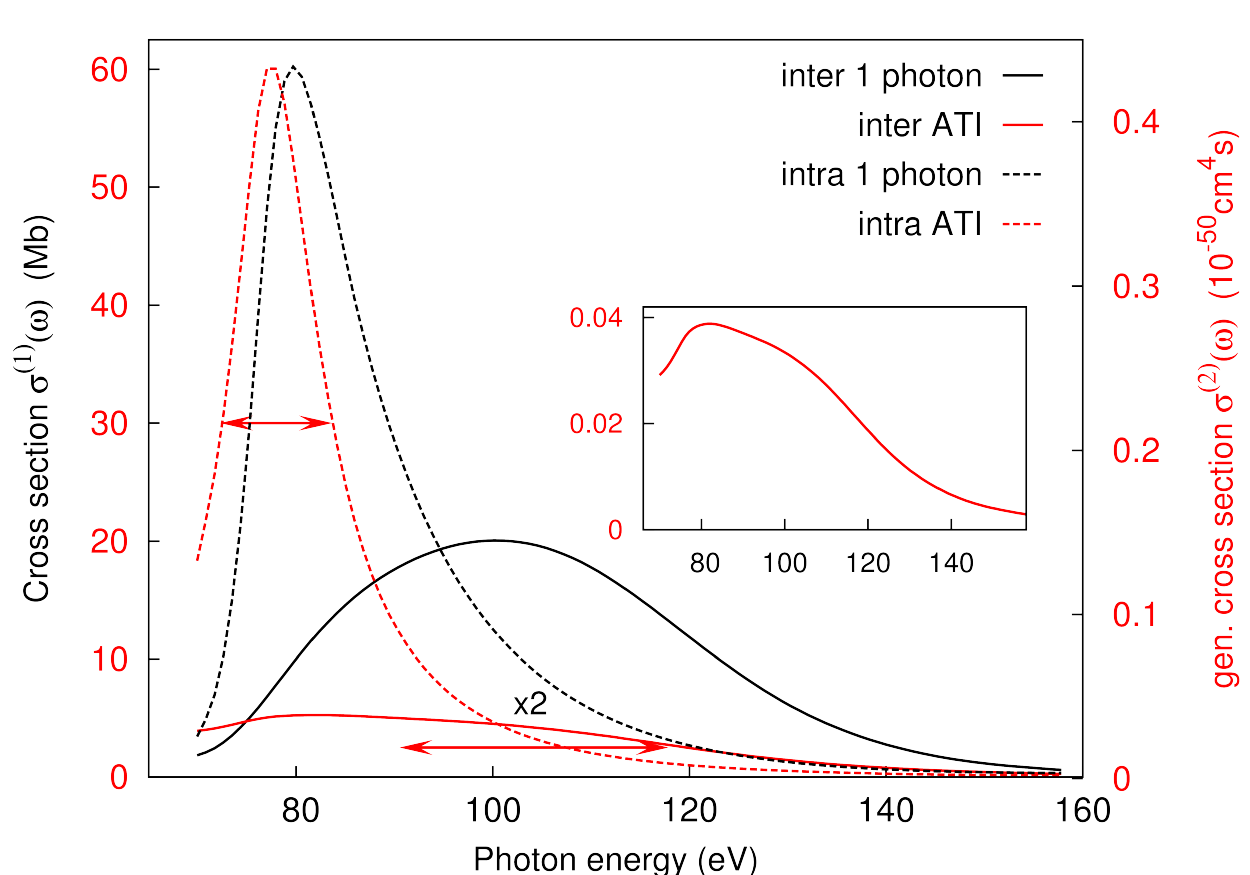}
 \caption{One-photon and \acs{ATI} (generalized) cross sections of xenon for the interchannel (solid lines) and intrachannel model (dashed lines) in the range of the \acs{GDR}. The interchannel \acs{ATI} cross section is much broader than the one-photon cross section, see the arrow spanning half the \acs{FWHM}. The inset magnifies the peculiar shape of the \acs{ATI} curve for the interchannel case.}
 \label{xeati:cross}
 \end{centering}
\end{figure}

\begin{figure}[!tb]
\centering
 \includegraphics[width=.7\linewidth]{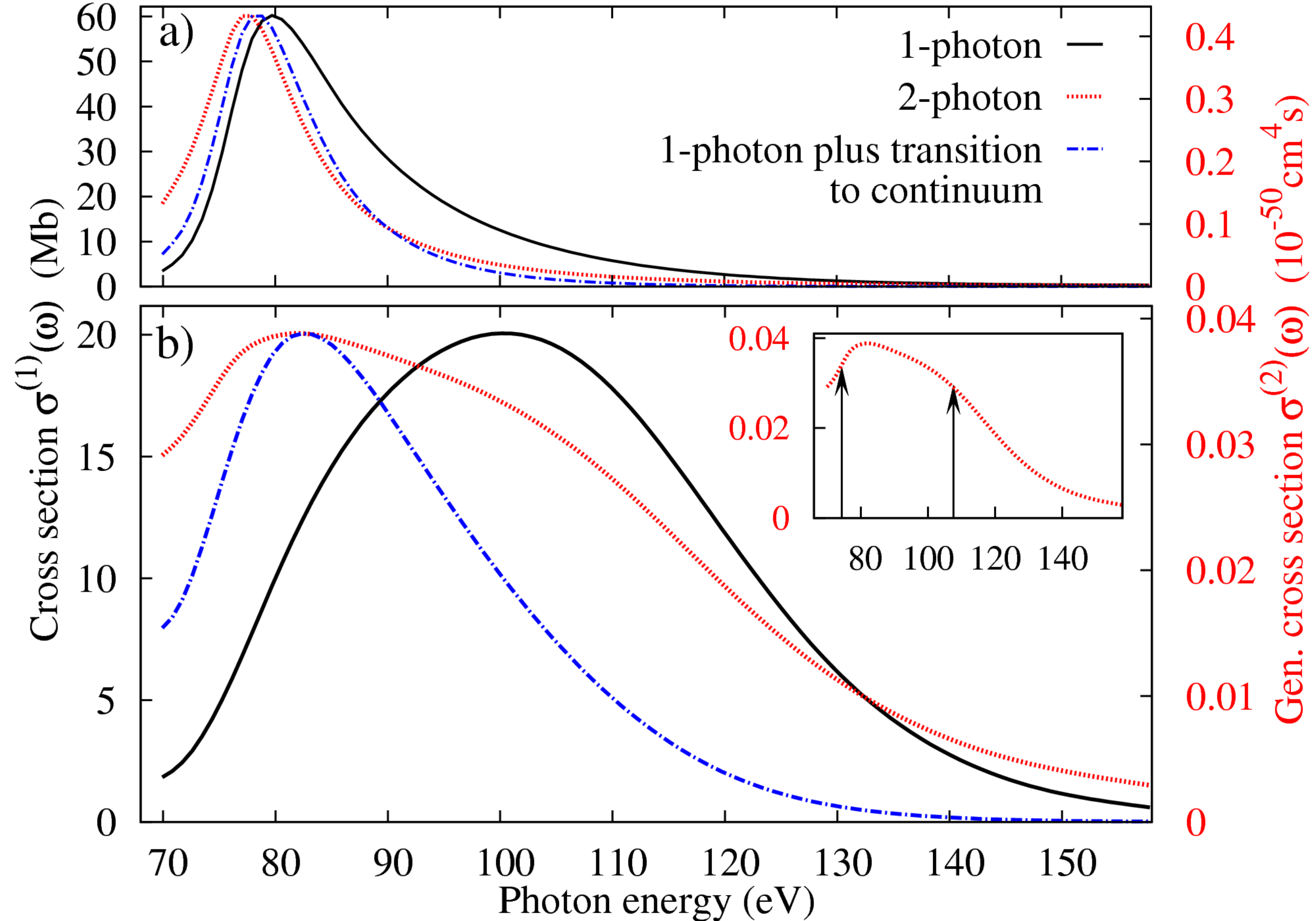}
 \caption{Photon-energy dependence of the calculated xenon cross sections: one-photon (solid black line) and generalized two-photon (dotted red line) cross sections within the intrachannel (panel a) and interchannel model (panel b)~\cite{Maz15}. The dash-dotted blue lines represent the result for the two-photon cross section within the two-step model with one single intermediate resonance state. The inset shows the full model two-photon cross section with two arrows indicating the energy position of the two underlying resonances~\cite{Che15}. Reprinted with permission from Ref.~\cite{Maz15}, \copyright ~Nature Publishing Group}
 \label{xeati:fig4}
\end{figure}

As a next step, the influence of collectivity on the cross sections is investigated over the whole photon-energy range of the \acs{GDR}. In Fig.~\ref{xeati:cross} the one-photon and generalized two-photon cross sections are plotted together for the intrachannel (dashed lines) and the interchannel model (solid lines). The two-photon cross section for the intrachannel model is red-shifted and narrower than the one-photon curve which is indicated by the red arrow spanning the \acs{FWHM}. This would also be a naive guess for the two-photon cross section, if we consider that the second photon induces a continuum-continuum transition. To understand this, let us assume that a single intermediate state, namely the \acs{GDR}, is populated resonantly in the two-step ionization process, see Fig.~\ref{xeati:fig1}~b). In perturbation theory, where the interaction of the atom with the light field, $\hat{H}_{\mathrm{int}}$, is treated as a perturbation, $\sigma^{(2)}$ can be factorized into two one-photon cross sections as long as the photon energy lies is the vicinity of a single, isolated one-photon resonance. The first photon excites the \acs{GDR}, the second photon ionizes the electron. Recalling Eq.~\eqref{transmatrixelement}, we obtain the two-photon cross section with the \acs{GDR} as one single intermediate state
\begin{equation}
 \sigma^{(2)}=\bigg| \frac{ \langle F | \hat{H}_{\mathrm{int}} | {\mathrm{GDR}} \rangle \langle {\mathrm{GDR}} | \hat{H}_{\mathrm{int}}| I \rangle}{E-E_{\mathrm{GDR}}+\frac{i}{2}\Gamma_{{\mathrm{GDR}}}+E_I}\bigg|^2 . \label{xeati:simpletwocross}
\end{equation} 
Now, the second photon initiates a continuum-continuum transition. The cross section of this transition follows a simple energy dependence of $E^{-l-7/2}$, cf. Sec.~$70$ of Ref.~\cite{bethe}, with $l$ being the angular momentum of the initial state. The \acs{GDR} exhibits mostly $f$-character, such that the exponent becomes $-13/2$. Therefore, for one single intermediate state the two-photon cross section is obtained by multiplying the one-photon ionization cross section with this energy-dependent factor. According to the two-step picture one expects a narrower two-photon peak that is shifted to lower energy, because the one-photon cross section for exciting an electron from the intermediate state into the continuum decreases monotonically with increasing energy. This is visualized in Fig.~\ref{xeati:fig4}~a). The intrachannel one-photon ionization cross section (black curve) multiplied by $E^{-13/2}$ is indeed shifted to smaller energies and exhibits also a smaller width, just like the two-photon cross section peak (red curve). Therefore, the behavior of the two-photon cross section in the intrachannel case can be qualitatively understood in terms of a sequential process involving the \acs{GDR} as a single intermediate state (dashed blue curve).

Coming back to Fig.~\ref{xeati:cross} we observe a fundamentally different behavior for the interchannel case: Unexpectedly, it predicts a significantly broader two-photon cross section curve (solid red curve) than for the one-photon case, demonstrated by the red arrow which spans only half the \acs{FWHM}. Moreover, the shape of the curve is peculiar, it does not resemble a simple Lorentzian profile, but exhibits a knee as can be seen more clearly in the inset of the figure. 
What is the origin of the broadening and the shape?

 To answer this question let us apply the two-step model to the interchannel scenario. As shown in Fig.~\ref{xeati:fig4}\,b) the resulting curve for the generalized two-photon absorption cross section within the simplified model (blue curve) is indeed shifted to a smaller energy. But the width is also decreased in strong contrast to the calculated two-photon cross section curve in the full model, which leads to the broadened red curve. Furthermore, the dashed blue curve underestimates the experimental cross section especially at $140$~eV by a considerable factor. This demonstrates that the simple model does not capture the physics of the full model if only a single resonance is taken into account as the intermediate state.
 
Besides its broad profile, the shape of the \acs{ATI} curve strongly suggests that there are two resonances underlying the broad \acs{GDR}, one being located at the peak of the curve and the other one near the knee-type structure. In the case of two resonances there must be a sum over the corresponding resonance states in the cross section expression~(\ref{xeati:simpletwocross})
\begin{equation}
 \sigma^{(2)}=\bigg|\sum_{M_{\mathrm {res}}}  \frac{ \langle F | \hat{H}_{\mathrm{int}} |M_ {\mathrm{res}} \rangle \langle M_ {\mathrm{res}} | \hat{H}_{\mathrm{int}}| I \rangle}{E-E_{M_\mathrm{res}}+\frac{i}{2}\Gamma_{M_{\mathrm{res}}}+E_I}\bigg|^2 . \label{xeati:crosseq}
\end{equation}
From this expression it becomes clear that once there are more than one intermediate state, interference terms between overlapping resonances arise, whose relative phase can result in a broadening and can change the shape of the cross section curve. Clearly, due to the cross terms in the \acs{ATI} cross section, the nonlinear process represents a more sensitive observable for testing the hypothesis of two resonances~\cite{Maz15}. 

The \acs{ATI} cross section provides the first hint to the fact that there exist two underlying resonances.
In order to investigate the resonance states further, the same strategy as in Sec.~\ref{arpack} is employed:
 In Ref.~\cite{Che15} the resonance states, which are similar to the tunneling states we encountered previously, are investigated in detail by diagonalizing the $N$-electron Hamiltonian. Similarly to our considerations above, the resonance states are exponentially divergent in the asymptotic region~\cite{sie}. Hence, it is difficult to access them by a Hermitian Hamiltonian. Again, in order to overcome this obstacle the Hamiltonian is rendered non-Hermitian; this time by the exterior complex scaling method (\acs{ECS})~\cite{nim} (instead of a \acs{CAP} as done above). Thereby the associated resonance wave functions are transformed into square-integrable functions.

The states of interest are accessible from the ground state by the absorption of one photon. Therefore, the relevant one-photon resonances are found by imposing the requirement that they possess a large overlap with the ground state coupled by one dipole step \cite{Che15}.
\begin{table}[!tb]
\begin{tabular}{|c|c|c|c|}
\hline
scheme  & resonance & Energy (eV) & ``lifetime'' (as)  \tabularnewline 
\specialrule{.2em}{.1em}{.1em} 

intrachannel &  $R_1,\ R_2,\ R_3$ &77 & 60 \tabularnewline
\hline
 \multirow{2}{*}{interchannel} &  $R_1$ &74.3  & 26 \tabularnewline
\cline{2-4} 
 & $R_2$ & 107.6 & 11\tabularnewline
\hline 
\end{tabular}\caption{Xenon resonances in the intra- and interchannel cases, data taken from Ref.~\cite{Che15}.}
\label{res:tab}
\end{table}
When doing so, three resonances are found in the intrachannel case, which form a group, cf. Tab.~\ref{res:tab}. They correspond to the three $4d_{\pm m}$, $|m|=0,1,2$, channels. The real part of the resonance group, $\approx 77$~eV, is consistent with the peak position and enhanced magnitude in the one-photon cross section calculated within the intrachannel model, cf. the black-dashed curve in Fig.~\ref{xeati:cross}. 
Activating the interchannel coupling, however, reveals two distinct underlying resonance states as shown in Tab.~\ref{res:tab}~\cite{Che15}. This means, that many-body correlations lead to resonances that cannot be attributed to single ionization pathways. The positions of the resonances, marked by arrows in the inset of Fig.~\ref{xeati:fig4} at $\approx 74$ and $108$~eV, are also consistent with our findings concerning the \acs{ATI} cross section. The wave function of the excited electron shows a prominent $f$-wave character for both resonances such that the xenon \acs{GDR} is dominated by $4d\rightarrow \epsilon f$ transitions. This legitimates our previous assumptions for the energy dependence in the two-step ionization model for the two-photon cross sections. More details can be found in Ref.~\cite{Che15}.

Summarizing, this example shows that the nonlinear response of a many-electron system to intense \acs{XUV} radiation can be used to unveil information about its collective behavior. The theoretical xenon \acs{ATI} cross section exhibits a knee-type structure which is not visible in the one-photon cross section. The agreement of the \acs{TDCIS} model with experimental results in the two-photon regime legitimates the prediction of two resonance states underlying the \acs{GDR}. 

\subsection{ATI in the x-ray regime}
We continue our investigation towards smaller wavelengths in the x-ray regime. Following the objective of Ref.~\cite{Til15} we ask how important \acs{ATI} is at hard x-ray photon energies. As described above, \acs{PES} are most adequate observables for quantifying the impact of \acs{ATI}. 

X-ray free-electron lasers (XFELs) provide ultrashort (hard) x-ray pulses at very high intensities. Due to their small wavelength x-rays provide high spatial resolution down to a few \r Angstr\"oms. Such pulses are of particular interest for the purposes of single-molecule imaging via coherent x-ray scattering at atomic resolution \cite{Chapman.Nature,Nature2000} but also for the investigation of electronic dynamics in atoms and (bio-) molecules on a time scale between attoseconds and tens of femtoseconds \cite{Nature420,kra}. 

Even though the interaction probability of x-rays with matter is low \cite{xdb2001}, in the high-intensity regime it might be necessary to consider the importance of nonlinear processes affecting electronic dynamics of atomic, molecular or solid-state target systems. 
With higher photon energy the probability for photoabsorption by electrons of the valence shell decreases significantly. In the x-ray regime the absorption probability for electrons in the valence shells at low intensities is negligible compared to visible or \acs{XUV} light. For instance, already at $1$\,keV a $2p$ electron in a carbon atom absorbs with a cross section of only $10^{-4}$~Mb. This cross section is $3$ orders of magnitude smaller than for the core electrons~\cite{yeh85}. The core electrons are more likely to absorb because they have a larger binding energy. When it comes to very high intensities inner-shell electrons might absorb even more than one photon despite of the high photon energy. Especially for imaging experiments at \acs{XFEL}s this circumstance could pose a problem because after diffraction from a sample the real space image needs to be retrieved from the image in the momentum-transfer space. For brighter illumination and a higher signal intensity higher pulse intensities are needed, e.g., in order to image the interior of a virus. One strategy to achieve significantly higher intensities at \acs{XFEL}s is to focus the pulses down to a few nanometers. \acs{ATI} at such high intensities might be a process that can produce considerable signal in the photon energy regime used in imaging experiments. 

Therefore, let us examine the role and the magnitude of nonlinear effects in the x-ray regime under high-intensity conditions that might become available soon at XFELs. The atomic potential is treated on the \acs{HFS} level, which, as discussed in Sec.~\ref{theory}, significantly reduces the computational cost as it spares the calculation of the exact Coulomb interaction between the electrons. The comparison with full \acs{TDCIS} calculations suggests that, indeed, in this photon-energy regime electron correlations are of minor importance~\cite{Til15}. 
Nevertheless, the investigation of the interaction of x-rays with atoms is computationally challenging. The grid size, the parameters for the calculation of the \acs{PES} using the wave-function splitting method and the propagation time step have to be adjusted in order to numerically capture the high-energy components, i.e., the fast oscillations, of the wave packet in time and space~\cite{Til15}. 

The applicability of the \acs{PES} calculations within \acs{TDCIS} in the x-ray regime is demonstrated for hydrogen by comparison with 
previous work \cite{PhysRevA.84.033425}, which shows nice agreement. 
Although under the conditions of short-wavelength x-rays our assumption of the 
dipole approximation (cf. Sec.~\ref{theory}) may no longer be valid, it is found that the results underestimate the nondipole results merely by a factor of $2-3$~\cite{Til15}. Nevertheless, going beyond the dipole approximation in the light-matter interaction would be highly desirable and remains an interesting topic of future investigations. Furthermore, it is found that it is appropriate to neglect the relativistic correction for the ionized electron.

\begin{table}[!tb]
\begin{tabular}{|c|c|c|c|c|}
\hline 
\multirow{2}{*}{E (keV)} & \multicolumn{4}{c|}{$K$-shell generalized two-photon cross sections (cm$^{4}$s)}\tabularnewline
\cline{2-5} 
 & Hydrogen & Carbon & Nitrogen & Oxygen \tabularnewline
\specialrule{.2em}{.1em}{.1em} 
8   & $1.44\times10^{-66}$ & $1.64\times10^{-62}$ & $3.21\times10^{-62}$ & $5.62\times10^{-62}$ \tabularnewline
\hline 
10 & $4.69\times10^{-67}$ & $4.61\times10^{-63}$ & $9.23\times10^{-63}$ & $1.70\times10^{-62}$  \tabularnewline
\hline 
12 & $1.72\times10^{-67}$ & $1.79\times10^{-63}$ & $3.82\times10^{-63}$ & $6.94\times10^{-63}$ \tabularnewline
\hline 
\end{tabular}\caption{Two-photon ATI cross sections for the $K$ shell of the light elements (E is the photon energy). Data reproduced from Ref.~\cite{Til15} with permission. All were calculated from integrating the corresponding photoelectron peaks at $10^{20}$~Wcm$^{-2}$ intensity and $0.12$~fs of pulse duration.}
\label{xray:tab}
\end{table}

Of course, as mentioned several times in this tutorial, in order to calculate a meaningful cross section we must make sure to be in the perturbative limit, i.e., the ionized populations remain very small. A Gaussian pulse of $10^{20}$~Wcm$^{-2}$ intensity and $0.12$~fs pulse duration is assumed at three typical hard x-ray photon energies of $8$, $10$, and $12$~keV. The cross sections are calculated by integrating the corresponding photoelectron peaks and dividing by the fluence, according to Eqs.~\eqref{fluences}. 
The two-photon $K$-shell \acs{ATI} cross sections for hydrogen and for the chemical elements that are commonly found in organic molecules, namely, carbon, nitrogen, and oxygen are presented in Table~\ref{xray:tab}~\cite{Til15}. We expect an increase in the \acs{ATI} cross section for higher $Z$ values because the electron binding energies are larger. This is indeed the case. Also, the cross section should decrease with larger photon energy because the ratio between the electron's binding energy and the photon energy decreases, and the electron appears almost to be free.\footnote{Recall that it is impossible for a free electron to absorb a photon, because energy and momentum cannot be conserved simultaneously.} 

\begin{figure*}[!b]
\includegraphics[width=.95\linewidth]{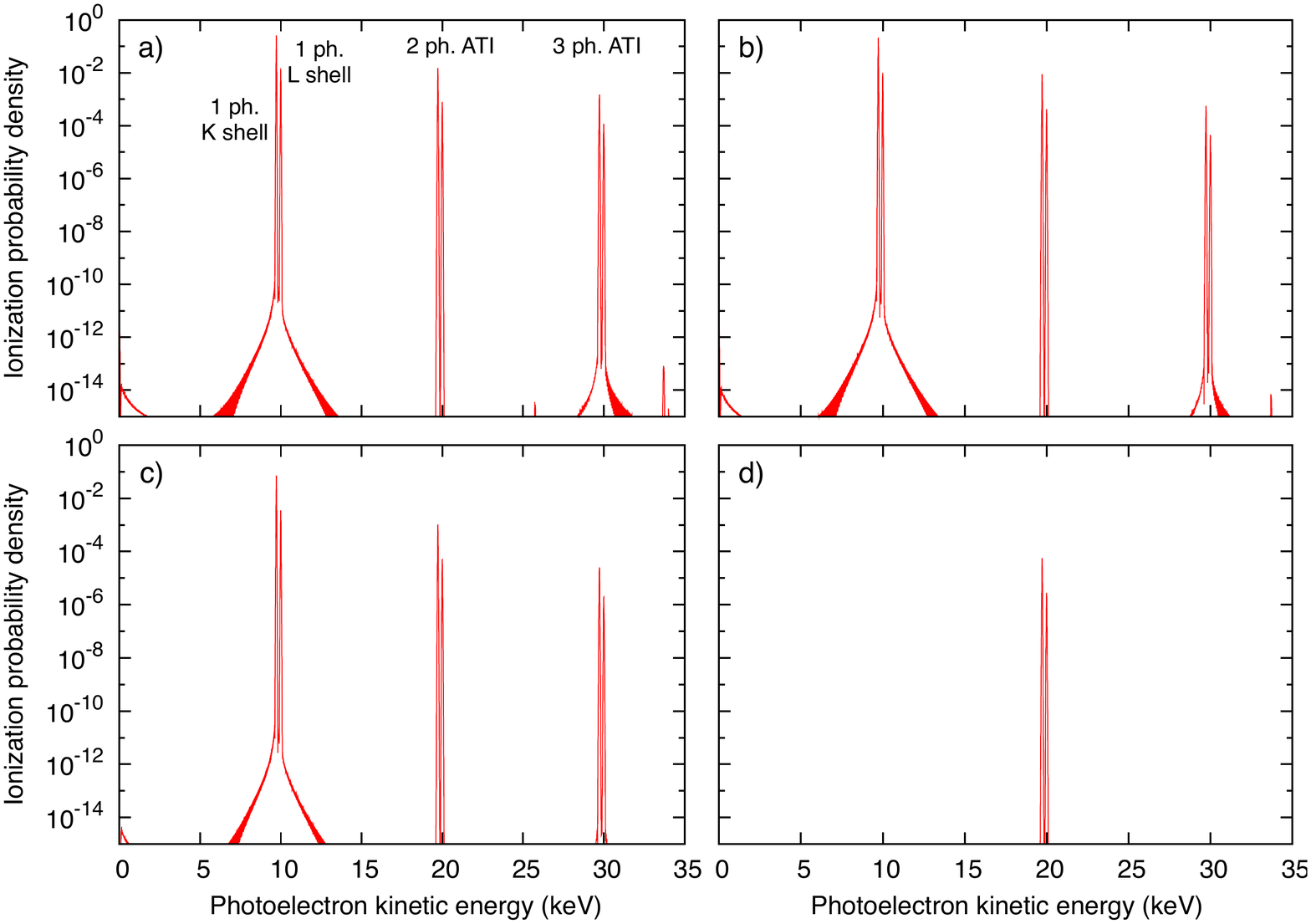}
\caption{\acs{PES} showing the one-photon ionization and the first two \acs{ATI} peaks of carbon for different angles~\cite{Til15}: a) $0$, b) $\pi/6$, c) $\pi/3$, d) $\pi/2$. The spectra are calculated for a pulse centered at $10$~keV photon energy, at an intensity of $10^{24}$~Wcm$^{-2}$ and $0.12$~fs pulse duration. Note that the height of the $K$-shell \acs{ATI} peak is greater than or comparable to the one-photon $L$-shell ionization peak. Reprinted with permission from Ref.~\cite{Til15}. \copyright IOP Publishing}
\label{fig.car}
\end{figure*}

Fig.~\ref{fig.car} shows the \acs{PES} of carbon at a photon energy of $10$~keV and a rather high intensity of $10^{24}$~Wcm$^{-2}$ for the angles $ \theta=0$, $\pi/6$, $\pi/3$, and $\pi/2$~\cite{Til15}.
The one-photon ionization peak and the first two \acs{ATI} peaks are shown. Each peak consists of two subpeaks, the one at lower energy being associated with ionization from the $K$ shell, the other one with $L$-shell ionization. 
At this intensity the depopulation due to $K$-shell \acs{ATI} in the direction $ \theta=0$ becomes comparable and even higher than valence one-photon ionization.

\begin{figure}[!tb] 
\begin{centering}
\includegraphics[width=.75\linewidth]{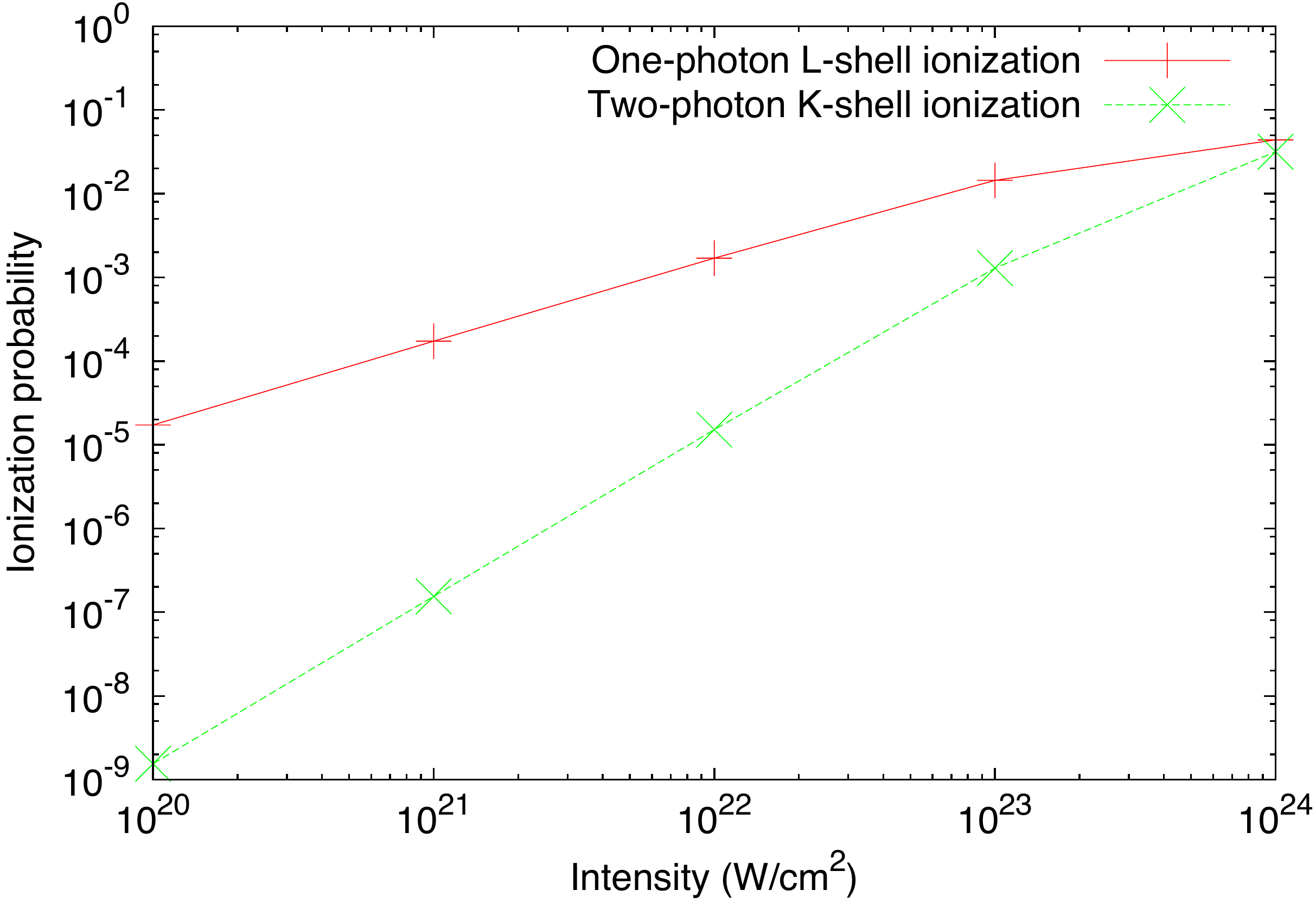}
\caption{Depopulations at different intensities due to two-photon \acs{ATI} and one-photon valence electron ionization for carbon, at $10$~keV photon energy and $0.48$~fs pulse duration~\cite{Til15}. The ionization probability is calculated by integrating the corresponding \acs{PES} peaks. Reprinted with permission from Ref.~\cite{Til15}. \copyright IOP Publishing}
\label{fig.int}
\end{centering}
\end{figure}

Next, the ionization probability of carbon for the cases of one-photon ionization out of the $L$ shell and of two-photon ionization out of the $K$ shell are presented for different pulse intensities. They are obtained by integrating the corresponding peaks over the energy and the angle as described in Ref.~\cite{Til15}. In Fig.~\ref{fig.int} the ionization probabilities are shown as a function of intensity on a double-logarithmic scale. We recognize the characteristic quadratic behavior of the two-photon \acs{ATI} peak and the linear behavior in the one-photon valence ionization probability from the slopes of the lines, remember that $\ln P_N\propto N\ln I $ from Eq.~\eqref{logdep}. Because of the high photon energy, deviations from the perturbative ionization do not play a role for the intensity regime below $10^{24}$~Wcm$^{-2}$. As mentioned above, between $10^{23}-10^{24}$~Wcm$^{-2}$ the ionization probability due to $K$-shell \acs{ATI} peak is comparable to the probability to ionize carbon with one photon out of the valence shells.

\begin{figure}[!tb] 
\includegraphics[width=.75\linewidth]{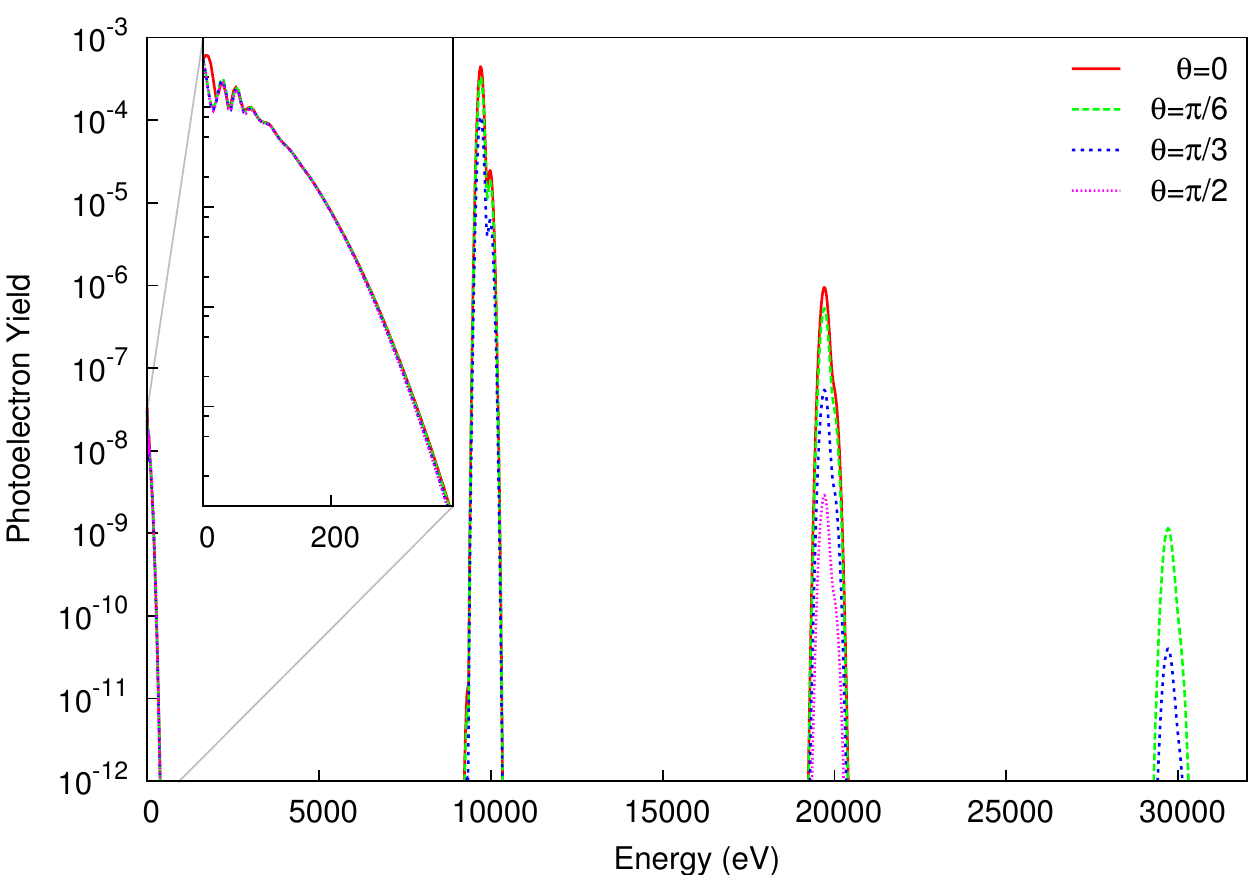}
\caption{Photoelectron yield for carbon, shown for a pulse of $12$~as duration and $3.5\cdot10^{22}$~Wcm$^{-2}$ intensity with a photon energy of $10$~keV after a sufficiently long propagation time~\cite{Til15}. The slow-electron peak (magnified in the inset) as well as the $1$-photon peak and the first two \acs{ATI} peaks are shown for $4$ different angles with respect to the polarization direction. Reprinted with permission from Ref.~\cite{Til15}. \copyright IOP Publishing}
\label{sep}
\end{figure}

Re-examining Fig.~\ref{fig.car}, a small peak around zero kinetic energy is observed, e.g., in the direction $\theta=0$. What is the origin of these slow electrons? As mentioned above, for the splitting method the final propagation time $T$ must be large enough to detect all slow electrons in the splitting region. Since the focus of the study lay on the fast electrons the propagation time was chosen shorter than necessary to collect all slow electrons. Therefore, the height of the peak shown in Fig.~\ref{fig.car} probably underestimates the electron yield. A new calculation for a pulse at $10$~keV photon energy, $12$~as pulse duration, and $3.5\cdot10^{22}$~Wcm$^{-2}$ intensity allows to observe the slow-electron peak in its full height. 
The resulting \acs{PES} are shown in Fig.~\ref{sep} for different angles. The height of the slow-electron peak is comparable to the two-photon \acs{ATI} peak in the direction of $\pi/2$. This suggests that the two processes are of the same, namely, of second order. We can understand this by realizing that the pulse, being short in the time domain, has a large bandwidth in the energy domain which leads to the energy width of the photoelectron peaks. For a pulse duration of $12$~as the energy bandwidth amounts to $\approx 210$~eV for the two-photon peak. When a valence electron absorbs one photon at a certain energy and re-emits a photon of slightly lower energy within the pulse bandwidth, slow electrons carrying away the excess energy are produced. Of course, the absorption and subsequent emission of one photon is also a second order process.

We conclude from this example that \acs{ATI} remains negligible for intensities at the most recent \acs{XFEL} experiments with hard x-rays. However, with photon energies at around $10$~keV or below, when entering the regime around $10^{23}$~Wcm$^{-2}$ the two-photon ionization probability of the core electrons reaches the same order of magnitude as one-photon valence ionization.

\section{\label{sec:6}Conclusion}

We studied the ionization of many-electron systems in various photon-energy regimes, spanning the infrared up to the hard x-ray regime within the \acs{TDCIS} scheme, focussing on the nonlinear response regime. Light-atom interactions at intensities high enough to induce multiphoton absorption processes were investigated and the information that photoelectron distributions provide was examined. In particular, xenon was studied as a prime example of an atomic system exhibiting strong correlation effects in the atomic shell. Using xenon as a model system it was shown that nonlinear spectroscopy can be employed as a tool to broaden our knowledge about the collective resonance behavior of atomic systems. With the help of \acs{TDCIS}, which captures electron correlation effects in the atomic shell, the substructure in the \acs{GDR} of xenon arising from collective effects in the $4d$ shell was uncovered. The underlying resonance states were characterized in detail by analyzing the eigenstates of the Hamiltonian. Furthermore, the diagonalization of the strong-field Hamiltonian in the tunneling regime revealed, that, strictly speaking, tunneling ionization is a nonadiabatic process and that in the case of few-cycle pulses the dynamics can be fully described by one single diabatic state. It was shown that \acs{ATI} in the x-ray regime does not play a crucial role at current \acs{FEL} intensities. However, when increasing the pulse fluence further nonlinear ionization might start to play a role.

Summarizing, the \acs{TDCIS} scheme has found many applications for the study of multiphoton ionization in the tunneling and perturbative multiphoton regimes employing also the calculation of photoelectron distributions. Particularly for the analysis of electron correlation effects it has proven beneficial and the applicability of the methods presented in this tutorial is by far not exhausted.

\section{Acknowledgment}
This work was supported by the Hamburg Centre of Ultrafast Imaging through the Louise-Johnson Fellowship.

\bibliographystyle{apsrev}
\bibliography{bibtutorial}

\end{document}